\documentclass[aps,prl,twocolumn,superscriptaddress,floatfix]{revtex4-2}
\usepackage{amssymb, amsmath, bm}
\usepackage{graphicx,onlyamsmath}

\usepackage{natbib}
\usepackage{xcolor}
\usepackage[normalem]{ulem}  

\begin{document}
\title{Many-particle hybridization of optical transitions from zero-mode Landau levels in HgTe quantum wells}

\author{S.~Ruffenach}
\thanks{These authors contributed equally to this work.}
\affiliation{Laboratoire Charles Coulomb (L2C), UMR 5221 CNRS-Universit\'{e} de Montpellier, F-34095 Montpellier, France}

\author{S.~S.~Krishtopenko}
\thanks{These authors contributed equally to this work.}
\affiliation{Laboratoire Charles Coulomb (L2C), UMR 5221 CNRS-Universit\'{e} de Montpellier, F-34095 Montpellier, France}

\author{A. V.~Ikonnikov}
\affiliation{Physics Department, M.V. Lomonosov Moscow State University, Moscow, 119991, Russia}

\author{C.~Consejo}
\affiliation{Laboratoire Charles Coulomb (L2C), UMR 5221 CNRS-Universit\'{e} de Montpellier, F-34095 Montpellier, France}

\author{J.~Torres}
\affiliation{Laboratoire Charles Coulomb (L2C), UMR 5221 CNRS-Universit\'{e} de Montpellier, F-34095 Montpellier, France}

\author{X.~Baudry}
\affiliation{CEA-LETI, MINATEC Campus, DOPT, F-38054 Grenoble, France}

\author{P.~Ballet}
\affiliation{CEA-LETI, MINATEC Campus, DOPT, F-38054 Grenoble, France}

\author{B. Jouault}
\affiliation{Laboratoire Charles Coulomb (L2C), UMR 5221 CNRS-Universit\'{e} de Montpellier, F-34095 Montpellier, France}

\author{F.~Teppe}
\email[]{frederic.teppe@umontpellier.fr}
\affiliation{Laboratoire Charles Coulomb (L2C), UMR 5221 CNRS-Universit\'{e} de Montpellier, F-34095 Montpellier, France}

\date{\today}

\begin{abstract}
We present far-infrared magnetospectroscopy measurements of a HgTe quantum well in the inverted band structure regime over the temperature range of 2 to 60~K. The particularly low electron concentration enables us to probe the temperature evolution of all four possible optical transitions originating from zero-mode Landau levels, which are split off from the edges of the electron-like and hole-like bands. By analyzing their resonance energies, we reveal an unambiguous breakdown of the single-particle picture indicating that the explanation of the anticrossing of zero-mode Landau levels in terms of bulk and interface inversion asymmetries is insufficient. Instead, the observed behavior of the optical transitions is well explained by their hybridization driven by electron-electron interaction. We emphasize that our proposed many-particle mechanism is intrinsic to HgTe quantum wells of any crystallographic orientation, including (110) and (111) wells, where bulk and interface inversion asymmetries do not induce the anticrossing of zero-mode Landau levels.
\end{abstract}

\maketitle

Despite significant progress over the past decade in the fabrication of atomically thin films of various materials~\cite{Q1,Q2,Q3,Q4,Q5}, quantum well (QW) heterostructures still remain the only widely available two-dimensional (2D) systems for studying phenomena related to the quantum spin Hall effect (QSHE)~\cite{Q6,Q7,Q8,Q9,Q10,Q11,Q12,Q13,Q14,Q15,Q16,Q17,Q18,Q19,Q20,Q21,Q22,Q23,Q24,Q25,Q26,Q27,Q28,Q29,Q30,Q31,Q32}. Among all the QW heterostructures, HgTe/CdHgTe QWs takes a special place, being the first system in which the QSHE was experimentally observed~\cite{Q7,Q8}. The key requirement for observing the QSHE in HgTe QWs is the inverted band ordering at the $\Gamma$ point of the Brillouin zone in wide QWs in which the first hole-like (\emph{H}$1$) subband lies above the lowest electron-like (\emph{E}$1$) one~\cite{Q6}. As the QW width $d$ decreases, the inverted gap between \emph{H}$1$ and \emph{E}$1$ subbands gradually closes until \emph{H}$1$ drops below \emph{E}$1$ at $d<d_c$, leading to a trivial band ordering of conventional semiconductor QWs. At critical QW width, $d=d_c$, HgTe QWs host a gapless state with massless Dirac fermions~\cite{Q33}. In addition to the QW width, the band ordering in HgTe QWs can also be changed by temperature~\cite{Q34,Q35,Q58,Q36}, hydrostatic pressure~\cite{Q36}, strain~\cite{Q37} or short-range disorder~\cite{Q38,Q39,Q40}.

\begin{figure}
\includegraphics [width=1.0\columnwidth, keepaspectratio] {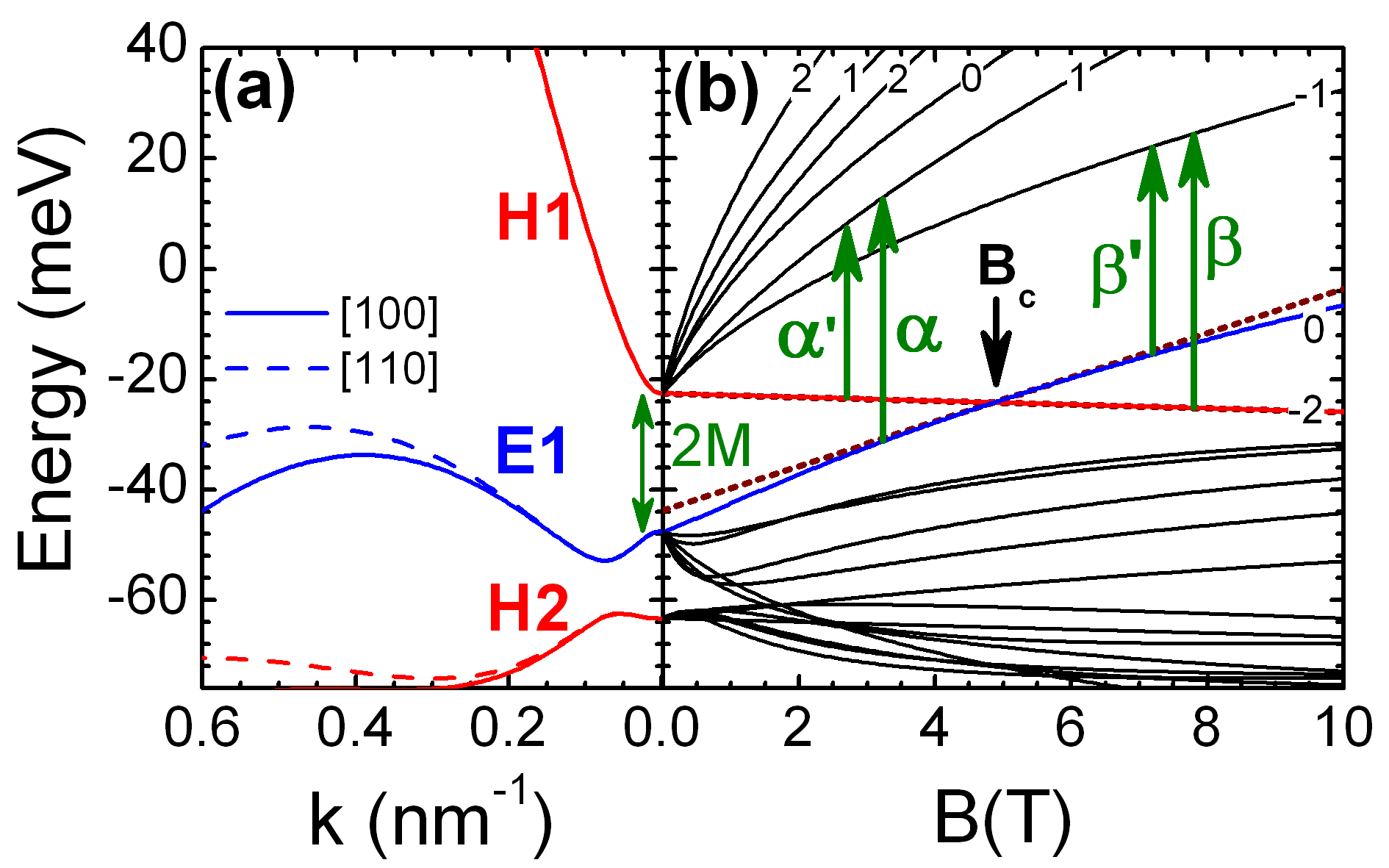} 
\caption{\label{Fig:1} \textbf{(a)} Band structure  in 8-nm-wide HgTe/Cd$_{0.7}$Hg$_{0.3}$Te QW at $T=2$~K grown on (001) CdTe buffer in the absence of the IIA effect. The blue and red curves represent band dispersion of electron-like (\emph{E}$1$) and hole-like (\emph{H}$1$ and \emph{H}$2$) subbands, calculated within the 8-band \textbf{k$\cdot$p} Hamiltonian~\cite{Q36}. The solid and dashed curves correspond to the quasimomentum orientation along the crystallographic directions (001) and (110), respectively. (\textbf{b)}~The energy of LLs in as a function of perpendicular magnetic field $B$. The numbers over the curves show the LL indices within the 8-band \textbf{k$\cdot$p} Hamiltonian~\cite{Q36}. The blue and red curves represent the zero-mode LLs from \emph{E}$1$ and \emph{H}$1$ subbands, respectively. The arrows represent the LL transitions observed in the vicinity of $B_c$~\cite{Q49,Q50,Q51,Q52,Q53,Q54,Q57}. The brown dotted curves represent the approximation of the linear dependence of the zero-mode LL energy on the magnetic field in the vicinity of $B_c$, used within the framework of Eq.~(\ref{eq:2}).}
\end{figure}

The inherent property of HgTe QWs, arising under perpendicular magnetic field $B$, is the presence of a particular pair of \emph{zero-mode} Landau levels (LLs), which split from the edges of \emph{E}$1$ and \emph{H}$1$ subbands. As these zero-mode LLs disperse in opposite directions with increasing $B$, they cross at the critical magnetic field $B_c$ under band inversion (see Fig.~\ref{Fig:1}), above which the inverted band ordering is transformed into the trivial one~\cite{Q7}. The lack of an inversion center caused either by the bulk inversion asymmetry (BIA)~\cite{Q41} or the interface inversion asymmetry (IIA)~\cite{Q42} leads to mixing of zero-mode LLs, resulting in their anticrossing in the vicinity of $B_c$. In HgTe QWs, it is impossible to experimentally distinguish between the BIA and IIA due to their similar contributions to the anticrossing gap. However, in bulk HgTe crystals, where IIA is naturally absent, the measurements of electric-dipole spin resonance did not revealed any BIA fingerprints~\cite{Q43}. The latter suggests that BIA-related effects should also be negligible in HgTe-based heterostructures as well. Therefore, in the following we assume that the anticrossing gap $\Delta$ of zero-mode LLs in HgTe QWs  represents the IIA strength of HgTe/CdHgTe heterojunctions.

At present, it is difficult to draw a definitive conclusion about the IIA strength in HgTe QWs, as the $\Delta$ values reported in the literature depends considerably on the experimental technique used to measure it. Particularly, magnetotransport~\cite{Q33,Q44,Q45,Q46} and photoconductivity~\cite{Q47} measurements performed using gated Hall bars show a crossing of zero-mode LLs within experimental accuracy, suggesting a weak IIA. This is consistent with the measurements of universal terahertz transparency, which reveal IIA-related energy splitting on the order of $0.6$~meV~\cite{Q48}. In contrast, far-infrared magnetospectroscopy reveals a fine structure in the LL transitions originating from zero-mode LLs~\cite{Q49,Q50,Q51,Q52,Q53,Q54}, indicating the presence of an anticrossing gap of about $5$~meV. This puzzling discrepancy in the reported $\Delta$ values suggests the presence of an unidentified contribution, which masks the real strength of IIA in HgTe QWs.

Recent investigations of the evolution of LL transitions from zero-mode LLs with varying electron concentration $n_S$ have revealed unexpectedly strong dependence of the energy gap at the $\Gamma$ point on $n_S$~\cite{Q57}. Such a strong dependence, where a 20-30\% percent change in the concentration results in a 50-70\% change in the energy gap~\cite{Q57}, has been attributed to the possible contribution from electron-electron (\emph{e-e}) interaction effects beyond the single-particle model.

In this work, by investigating the temperature evolution of all four possible LL transitions from the zero-mode LLs in HgTe QW with very low electron concentration, we reveal another unambiguous evidence of the breakdown of single-particle picture based on IIA. By analyzing the difference in resonance energies, we demonstrate that the observed anticrossing behavior in the vicinity of $B_c$ is well explained by the hybridization of LL transitions induced by \emph{e-e} interactions \emph{even in the absence of IIA}. This indeed indicates a weak IIA strength, consistent with previous results obtained by magnetotransport~\cite{Q33,Q44,Q45,Q46}, photoconductivity~\cite{Q47} and terahertz transparency measurements~\cite{Q48}. Importantly, the proposed many-particle mechanism that accounts for the observed behavior of LL transitions in the vicinity of $B_c$ is
intrinsic to HgTe QWs of arbitrary orientation, including (110) and (111) QWs, where the presence of IIA does not induce anticrossing of the zero-mode LLs.

The 8-nm-wide HgTe/Cd$_{0.7}$Hg$_{0.3}$Te QW studied in this work was grown in CEA-LETI by molecular beam epitaxy (MBE) on a (001)-oriented CdTe substrate with a relaxed CdTe buffer~\cite{Q59}. Figure~\ref{Fig:1} shows the band structure and the energy of LLs as a function of $B$ at $T=2$~K as expected for our sample in the absence of IIA. The calculations were performed using the 8-band \textbf{k$\cdot$p} Hamiltonian~\cite{Q36}, assuming a symmetric QW profile, which eliminates structure inversion asymmetry (SIA). To calculate the LLs, we apply the axial approximation by omitting the warping terms in the Hamiltonian~\cite{Q36}. The pair of LLs with indices $N=-2$ and $N=0$ identified as ``zero-mode'' LLs~\cite{Q7,Q8,Q49} is marked in red and blue, respectively. Details of the calculations and LL notation within the 8-band \textbf{k$\cdot$p} Hamiltonian can be found in Ref.~\cite{Q36}.

In the absence of IIA, only two transitions from the zero-mode LLs are allowed, following the conventional selection rule $\Delta N=\pm1$ imposed by angular momentum conservation~\cite{Q49}. These LL transitions are marked in Fig.~\ref{Fig:1} as $\alpha$ and $\beta$ according to the notation of Schultz~\emph{et~al.}~\cite{Q60}. In contrast, the $\alpha'$ and $\beta'$ transitions from the zero-mode LLs both correspond to ``spin-flip'' transitions~\cite{Q7,Q8}, which are forbidden in the single-particle picture in the absence of IIA. The presence of IIA mixes the zero-mode LLs with opposite spins, enabling the observation of $\alpha'$ and $\beta'$ transitions in the vicinity of $B_c$.

Importantly, in the single-particle picture, differences in the resonance energies $\hbar\omega_{\alpha'}-\hbar\omega_{\alpha}$ and $\hbar\omega_{\beta'}-\hbar\omega_{\beta}$ for the transition pairs ($\alpha$,~$\alpha'$) and ($\beta$,~$\beta'$) directly determine the energy splitting between the zero-mode LLs, $\epsilon^{(+)}_0$ and $\epsilon^{(-)}_0$:
\begin{equation}
\label{eq:1}
\Delta E=\left|\hbar\omega_{\alpha'}-\hbar\omega_{\alpha}\right|=\left|\hbar\omega_{\beta'}-\hbar\omega_{\beta}\right|=
\left|\epsilon^{(+)}_0-\epsilon^{(-)}_0\right|.
\end{equation}
This can be analytically calculated within the low-energy Dirac-like model proposed by Bernevig~\emph{et~al.}~\cite{Q6}:
\begin{equation}
\label{eq:2}
\Delta E=\sqrt{(2M)^2\left(1-\dfrac{B}{B_c}\right)^2+\Delta^2},
\end{equation}
where $M$ is the mass parameter that determines the energy gap at the $\Gamma$ point~\cite{Q57}. The same result can be also derived for (001) HgTe QWs within the multi-band \textbf{k$\cdot$p} Hamiltonian, assuming a linear magnetic-field-dependence of $\epsilon^{(+)}_0$ and $\epsilon^{(-)}_0$ in the vicinity of $B_c$ in the absence of IIA~\cite{SM}. In this case, the values of $2|M|$ in Eq.~(\ref{eq:2}) are always smaller than the real energy gap at the $\Gamma$ point (see Fig.~\ref{Fig:1}).

\begin{figure*}
\includegraphics [width=2.0\columnwidth, keepaspectratio] {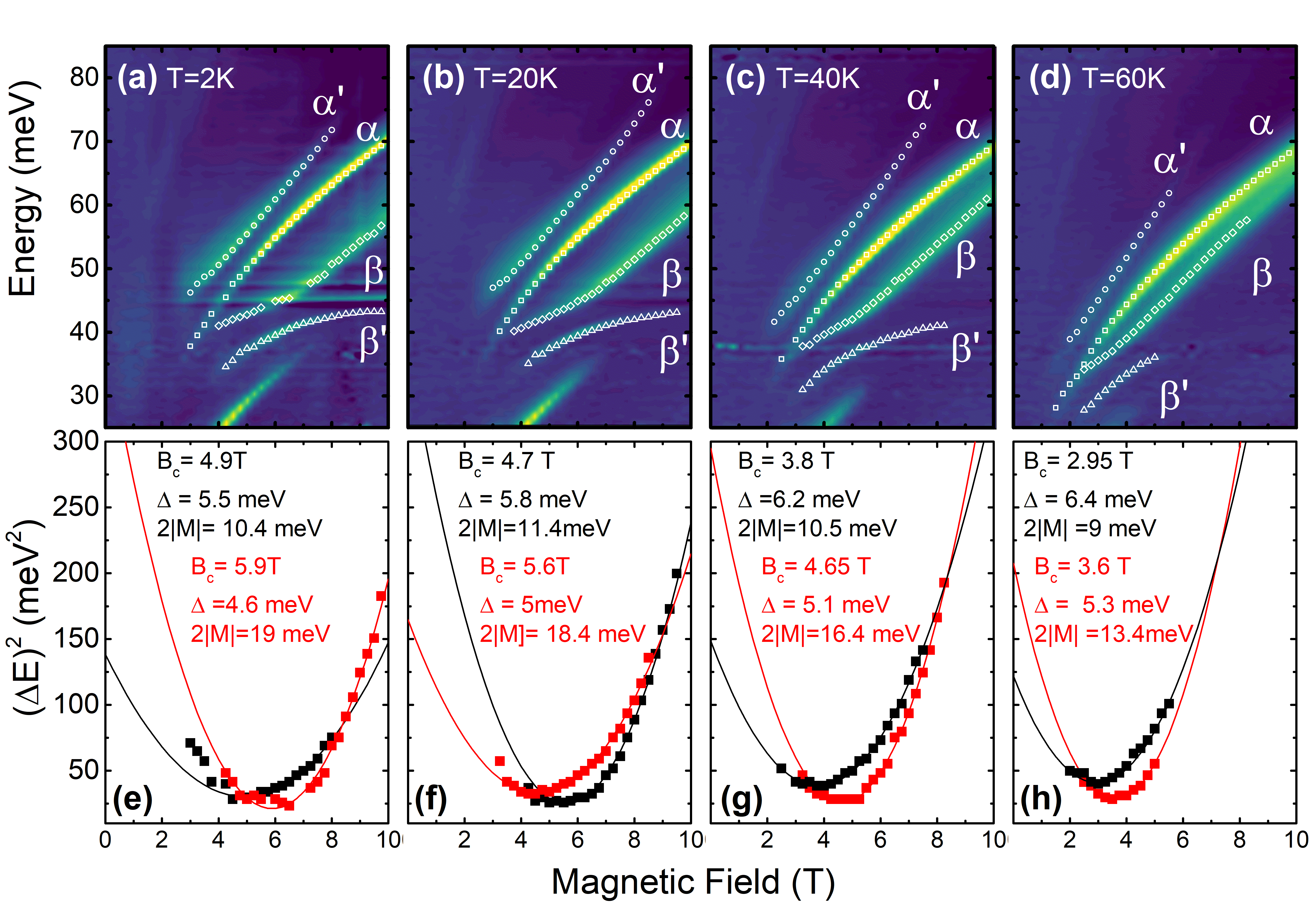} 
\caption{\label{Fig:2} \textbf{(a)-(d)} Color maps of magnetoabsorption showing $\alpha$, $\alpha'$, $\beta$, $\beta'$ LL transitions as a function of magnetic field, measured at different temperatures $T$: (a)~$2.0$~K, (b)~$20$~K, (c)~$40$~K, (d)~$60$~K. The symbols represent position of the magnetoabsorption lines, whose energies are used in the evaluation of $\Delta{E}$. \textbf{(e)-(h)} Square of the energy difference for $|\hbar\omega_{\alpha'}-\hbar\omega_{\alpha}|$ and $|\hbar\omega_{\beta'}-\hbar\omega_{\beta}|$ at the same temperatures as in the respective top panels. The black and red solid curves are the fitting to Eq.~(\ref{eq:2}) for the pairs ($\alpha$,~$\alpha'$) and ($\beta$,~$\beta'$), respectively. The results for other temperatures are provided in the Supplemental Material~\cite{SM}.}
\end{figure*}

Figure~\ref{Fig:2}(a-d) presents the magnetoabsorption spectra of our sample measured in the Faraday configuration at different temperatures by using a Fourier-transform spectrometer coupled to a 16-T superconducting coil~\cite{Q52,Q58}. All spectra were normalized by the sample transmission at zero magnetic field. Since we are interested in the LL transitions from zero-mode LLs in the vicinity of $B_c$, we present only the high-energy parts of the spectra with the traces of $\alpha$, $\alpha'$, $\beta$, $\beta'$ transitions, which look qualitatively similar to those reported in Ref.~\cite{Q57}. Although we did not have the opportunity to perform magnetotransport measurements during the temperature-dependent magnetospectroscopy, a clear conclusion about the low concentration in our sample can be drawn based on the presence of $\beta$ transition at all temperatures.

Indeed, a natural requirement for observing the $\beta$ transition is that the filling factor $\nu$ of the LLs in the conduction band remains significantly lower than two, which guarantees the presence of available states in the LL with $N=-1$ (see Fig.~\ref{Fig:1}). Therefore, the observation of the $\beta$ transition at $B\simeq2.0$~T and $T=60$~K (see Fig.~\ref{Fig:2}(d)) indicates that $n_S$ is significantly below $1.0\cdot10^{11}$~cm$^{-2}$. Such low concentrations, which were not accessible in the previous studies~\cite{Q57}, make it possible to fit the difference in resonant energies by Eq.~(\ref{eq:2}) for both ($\alpha$,~$\alpha'$) and ($\beta$,~$\beta'$) pairs of the LL transitions in the whole temperature range -- see Fig.~\ref{Fig:2}(e-h).

\begin{figure}
\includegraphics [width=1.0\columnwidth, keepaspectratio] {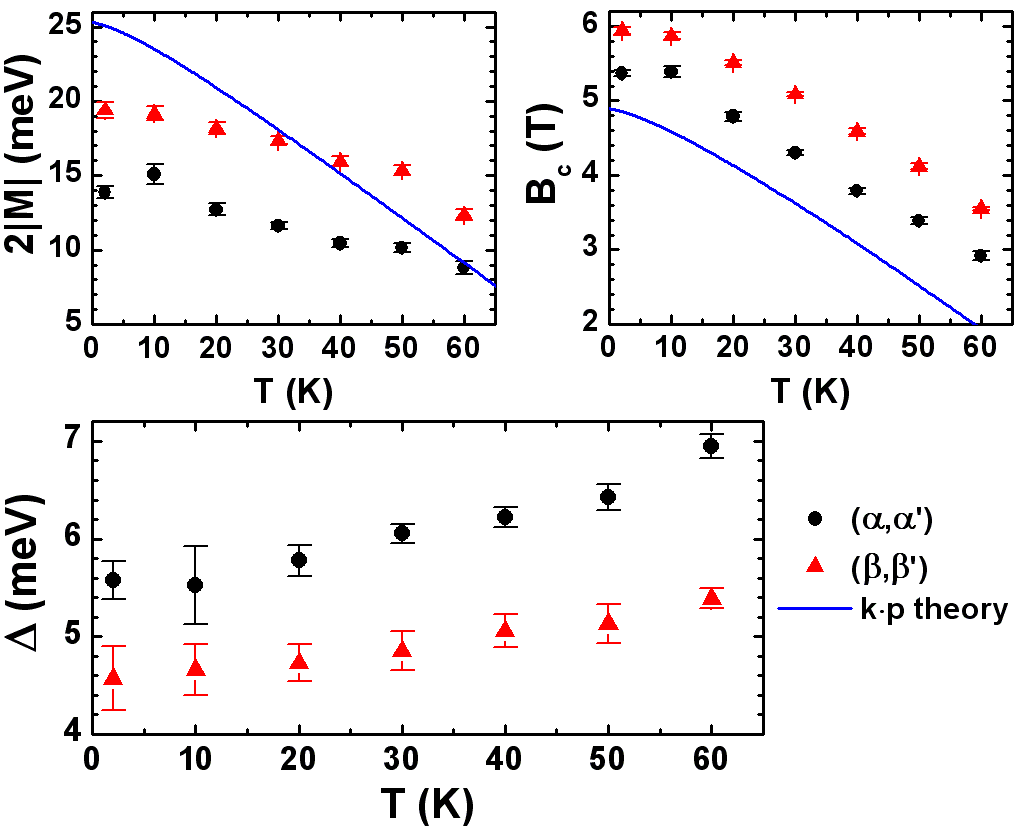} 
\caption{\label{Fig:3} Temperature evolution of $|M|$, $B_c$ and $\Delta$ extracted by fitting the difference in resonant energies for the pairs of ($\alpha$,~$\alpha'$) LL transitions (black circles) and ($\beta$,~$\beta'$) LL transitions (red triangles) by Eq.~(\ref{eq:2}). The blue curves represent the calculations of $|M|$ and $B_c$ performed by using the 8-band \textbf{k$\cdot$p} Hamiltonian~\cite{Q36}.}
\end{figure}

Figure~\ref{Fig:3} summarizes the values of $\Delta$, $B_c$ and $M$ as a function of temperature for both pairs of the transitions. It is clearly seen that all parameters $\Delta$, $B_c$ and $M$ extracted from $|\hbar\omega_{\alpha'}-\hbar\omega_{\alpha}|$ and $|\hbar\omega_{\beta'}-\hbar\omega_{\beta}|$ differ significantly from each other \emph{at all temperatures}. These differences cannot be explained in principle within the single-particle framework and thus indicate a breakdown of the single-particle picture. For comparison, we also provide the temperature dependence of $M(T)$ and $B_c(T)$ calculated using the 8-band \textbf{k$\cdot$p} Hamiltonian in the absence of IIA~\cite{Q36}. We emphasize in particular that the IIA does not lead to a renormalization of $M$ and $B_c$ in the (001) HgTe QW and can therefore be neglected. In contrast, for QWs grown along other crystallographic directions IIA must be taken into account when calculating $B_c$ and $M$~\cite{SM}.

As seen in Fig.~\ref{Fig:3}, the fitting parameters qualitatively reproduce the decreasing temperature dependence of $M$ and $B_c$, which is associated with the topological phase transition occurring at high temperatures~\cite{Q34,Q35,Q36}. Note that in the IIA-based single-particle picture, the theoretical $M(T)$ dependence is expected to always exceed the fitted values due to the nonlinear behavior of the zero-mode LL energy away from $B_c$ (see Fig.~\ref{Fig:1}). In turn, there are no any restrictions on the $B_c$ values obtained by using Eq.~(\ref{eq:2}). Therefore, given that theoretical calculations based on the 8-band \textbf{k$\cdot$p} Hamiltonian were previously in very good agreement with experimental $B_c$ values obtained from magnetotransport measurements~\cite{Q45}, the discrepancy observed in Fig.~\ref{Fig:3} can also be attributed to a breakdown of the single-particle picture used to interpret the LL magnetospectroscopy results.

Let us now discuss the mechanism beyond the single-particle picture that leads to the difference in transition energies $|\hbar\omega_{\alpha'}-\hbar\omega_{\alpha}|$ and $|\hbar\omega_{\beta'}-\hbar\omega_{\beta}|$ in the vicinity of $B_c$, which are nevertheless well fitted by Eq.~(\ref{eq:2}). It has long been understood that each inter-LL excitations in the quantum Hall regime, where LLs are well separated, can be considered as neutral collective modes described in terms of \emph{magnetic excitons} (MEs)~\cite{Q61,Q62,Q62b,Q62c} -- i.e., a bound state of a hole in a filled LL and an electron in an empty one. The long-wavelength limit of certain MEs, such as magnetoplasmons, defines the resonant energy of magnetoabsorption lines, which is known to be insensitive to \emph{e-e} interaction in 2D systems with parabolic band dispersion~\cite{Q63}. The band nonparabolicity, which naturally arises in HgTe QWs, not only leads to the emergence of multiple magnetoplasmonic modes with different energies, but also enables interactions between such MEs in the long-wave limit~\cite{Q64,Q65,Q66,Q67,Q68,Q69}.

We further show that the observed anticrossing behavior in the vicinity $B_c$ can be qualitatively explained by the influence of \emph{e-e} interaction on the hybridization of LL transitions \emph{even in the absence of IIA}. For simplicity, we focus on the range of magnetic fields in the vicinity of $B_c$, which allows us to treat the ($\alpha$,~$\alpha'$) and ($\beta$,~$\beta'$) transition pairs as being separated in energy from each other and from other LL transitions. In this case, to describe the hybridization of the $\alpha$ and $\alpha'$ transitions due to \emph{e-e} interaction, we can therefore restrict our analysis to these two basis states only. As a result the effective ME Hamiltonian takes the form~\cite{SM}:
\begin{equation}
\label{eq:3}
\hat{H}_{\mathrm{ME}\alpha\alpha'}=
\begin{pmatrix}
\hbar\omega_{\alpha}^{(0)}+\delta_{\alpha\alpha}^{(e-e)} & \left(\nu_{0}-\nu_{1}\right)\Delta_{\alpha\alpha'}^{(e-e)}/2 \\
\left(\nu_{-2}-\nu_{1}\right)\Delta_{\alpha\alpha'}^{(e-e)}/2 & \hbar\omega_{\alpha'}^{(0)}+\delta_{\alpha'\alpha'}^{(e-e)}
\end{pmatrix},
\end{equation}
where $\hbar\omega_{\alpha}^{(0)}$ and $\hbar\omega_{\alpha'}^{(0)}$ are the resonant energies defined by the LLs in the single-particle approximation, $\nu_{N}$ is the filling factor of each LL involved in $\alpha$ and $\alpha'$ transitions (see Fig.~\ref{Fig:1}), while $\delta_{\alpha\alpha}^{(e-e)}$, $\delta_{\alpha'\alpha'}^{(e-e)}$ and $\Delta_{\alpha\alpha'}^{(e-e)}$ being functions of $n_S$ and $B$~\cite{SM} describe the contribution from the \emph{e-e} interaction. The diagonal terms $\delta_{\alpha\alpha}^{(e-e)}$ and $\delta_{\alpha'\alpha'}^{(e-e)}$ change the energies of the $\alpha$ and $\alpha'$ excitons from their single-particle values. These terms include the exciton binding energy, as well as the interaction energy between the exciton and the electrons below the Fermi level~\cite{SM}. In turn, the off-diagonal term $\Delta_{\alpha\alpha'}^{(e-e)}$ describes the hybridization of two excitons, caused by the interaction between electrons and holes forming the $\alpha$ and $\alpha'$ excitons. Due to this hybridization, the $\alpha'$ exciton, being ``dark'' within the single-particle picture in the absence of IIA, contributes to the magnetoabsorption when the \emph{e-e} interaction is taken into account.

Let us now demonstrate that the ME picture also formally accounts for the good agreement between the fitted energy differences and Eq.~(\ref{eq:2}) as shown in Fig.~\ref{Fig:2}(e-h). Knowing the eigenvalues $\hbar\omega_{\alpha}$ and $\hbar\omega_{\alpha'}$ of $\hat{H}_{\mathrm{ME}\alpha\alpha'}$, the difference in the resonant energies of $\alpha$ and $\alpha'$ transitions is written as
\begin{multline}
\label{eq:4}
\left(\hbar\omega_{\alpha'}-\hbar\omega_{\alpha}\right)^2=
\left[\hbar\omega_{\alpha}^{(0)}-\hbar\omega_{\alpha'}^{(0)}+
\delta_{\alpha\alpha}^{(e-e)}-\delta_{\alpha'\alpha'}^{(e-e)} \right]^2+\\~
\left(\nu_{-2}-\nu_{1}\right)\left(\nu_{0}-\nu_{1}\right){\Delta_{\alpha\alpha'}^{(e-e)}}^2.
\end{multline}
By approximating the expression in square brackets as linear in $B$ near $B_c$, and assuming that
$\Delta=\sqrt{\left(\nu_{-2}-\nu_{1}\right)\left(\nu_{0}-\nu_{1}\right)}{\Delta_{\alpha\alpha'}^{(e-e)}}$ varies weakly with the magnetic field, Eq.~(\ref{eq:4}) can easily be reduced to the form of Eq.~(\ref{eq:2}). We note that small systematic deviations from Eq.~(\ref{eq:2}) appear in the magnetic-field range, in which the energies of the $\alpha$ and $\beta$ transitions approach each other. In this regime, the \emph{e-e} interaction induces the many-particle hybridization of these transitions as well~\cite{SM}. As a result, the transition pairs ($\alpha$, $\alpha'$) and ($\beta$, $\beta'$) can no longer be considered as fully independent, resulting in the deviation of $\left(\hbar\omega_{\alpha'}-\hbar\omega_{\alpha}\right)^2$ and $\left(\hbar\omega_{\beta'}-\hbar\omega_{\beta}\right)^2$ from Eq.~(\ref{eq:2}).

Importantly, the ME picture naturally explains the different values of the fitting parameters obtained in the analysis of the transition pairs ($\alpha$,~$\alpha'$) and ($\beta$,~$\beta'$). Indeed, by deriving the effective Hamiltonian $\hat{H}_{\mathrm{ME}\beta\beta'}$ for the $\beta$ and $\beta'$ MEs, one can verify that the many-particle contribution to their energies $\hbar\omega_{\beta}^{(0)}$ and $\hbar\omega_{\beta'}^{(0)}$ is determined by the matrix elements of the \emph{e-e} interaction, which are different from those for $\hbar\omega_{\alpha}^{(0)}$, $\hbar\omega_{\alpha'}^{(0)}$ and $\Delta_{\alpha\alpha'}^{(e-e)}$~\cite{SM}.

In turn, the temperature dependence of $\Delta$ shown in Fig.~\ref{Fig:3}, extracted from the analysis of the ($\alpha$,$\alpha'$) and ($\beta$,~$\beta'$) pairs, can also be qualitatively explained within the ME picture, assuming that the LL filling factor $\nu$ in the conduction band in the vicinity of $B_c$ remains below \emph{unity at all temperatures}. The latter gives $n_S\simeq5.0\cdot10^{10}$~cm$^{-2}$, which is consistent with the earlier estimate based on the observation of the $\beta$-transition at $60$~K. In the case of $\nu<1$, $\nu_{1}=\nu_{-1}=0$ (see Fig.~\ref{Fig:1}), and both hybridization energies $\sqrt{\nu_{-2}\nu_{0}}\Delta_{\alpha\alpha'}^{(e-e)}$ and $\sqrt{\nu_{-2}\nu_{0}}\Delta_{\beta\beta'}^{(e-e)}$ have the same dependence on the filling factor~\cite{SM}:
\begin{equation}
\label{eq:5}
\sqrt{\nu_{-2}\nu_{0}}\Delta_{\alpha\alpha'}^{(e-e)},\sqrt{\nu_{-2}\nu_{0}}\Delta_{\beta\beta'}^{(e-e)}\sim\sqrt{\nu}\sim\sqrt{\dfrac{n_S}{B_c}}.
\end{equation}
Taking into account the theoretical dependence of $B_c(T)$ shown in Fig.~\ref{Fig:3} and, for simplicity, assuming a weak temperature dependence of $n_S$, 
Eq.~(\ref{eq:5}) yields a factor of $1.58$ between $60$~K and $2$~K. Experimentally, the corresponding ratios extracted from Fig.~\ref{Fig:3} are $(1.25\pm0.06)$ and $(1.18\pm0.06)$ for the ($\alpha$,~$\alpha'$) and ($\beta$,~$\beta'$) pairs, respectively. The remaining discrepancy can be attributed to the unaccounted magnetic-field dependence of the \emph{e-e} interaction matrix elements~\cite{SM}, as well as to possible temperature-induced redistribution of carriers between the zero-mode LLs and hole-like LLs from the valence band.

\begin{figure}
\includegraphics [width=1.0\columnwidth, keepaspectratio] {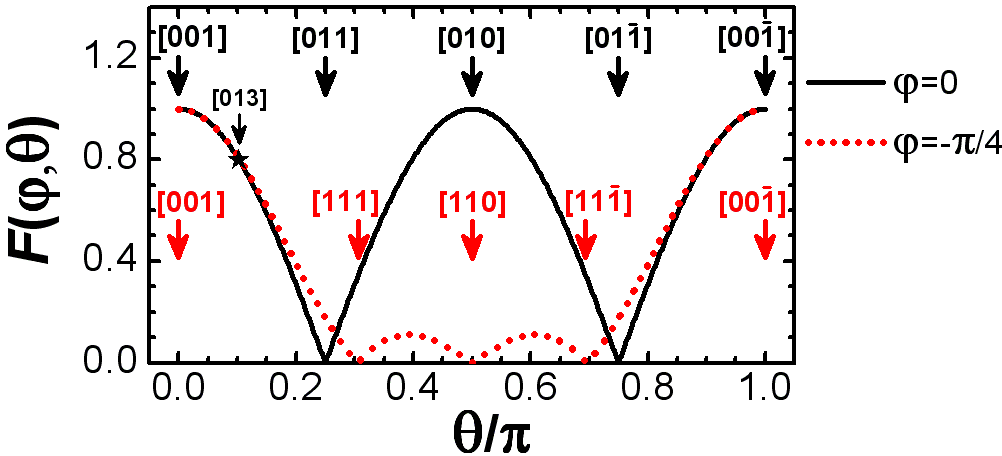} 
\caption{\label{Fig:4} Function ${F}(\varphi,\theta)$ that defines the IIA-induced anticrossing gap between zero-mode LLs within the single-particle picture. The arrows indicate the $\theta$ angles corresponding to specific growth directions.}
\end{figure}

Finally, we note that the proposed ME mechanism for the observed anticrossing behavior of LL transitions is universal and should be intrinsic to HgTe QWs of arbitrary crystallographic orientations. In contrast, the anticrossing gap between zero-mode LLs, used to explain the observed transition behavior in the vicinity of $B_c$ within the single-particle picture, strongly depends on the QW growth direction. Indeed, one can demonstrate that the IIA-induced anticrossing gap in HgTe QW of arbitrary orientation takes the form~\cite{SM}:
\begin{equation}
\label{eq:6}
\Delta(\varphi,\theta)=\Delta_0{F}(\varphi,\theta),
\end{equation}
where the angles $\varphi$ and $\theta$ defines the growth direction of HgTe QW with respect to the main crystallographic axes, $\Delta_0$ is a factor that has weak dependence on the QW growth orientation, and ${F}(\varphi,\theta)$ is written as~\cite{SM}
\begin{equation}
\label{eq:7}
{F}^2=\cos^2{2\theta}\cos^2{2\varphi}
+\dfrac{\cos^2{\theta}\left(3\cos^2{\theta}-1\right)^2\sin^2{2\varphi}}{4}
\end{equation}
The dependence of ${F}(\varphi,\theta)$ on $\theta$ for several values of $\varphi$ is shown in Fig.~\ref{Fig:4}. This confirms that the IIA-induced anticrossing of zero-mode LLs is absent in symmetric (110) and (111) HgTe QWs (as well as in equivalent ones). Interestingly, the same conclusion was previously made for the BIA-induced anticrossing gap~\cite{Q51}. Thus, in QWs of these orientations, the observation of anticrossing behaviour of LL transitions in the vicinity $B_c$ can only be explained within the many-particle ME framework.

In summary, by investigating the temperature evolution of all four possible LL transitions from the zero-mode LLs in HgTe QW with very low electron concentration, we reveal an unambiguous breakdown of the single-particle picture that has been widely used previously to describe the behavior of their resonant energies in a magnetic field. Alternatively, we show that the observed anticrossing behavior of these transitions is well explained by their hybridization driven by the \emph{e-e} interactions within the many-particle ME picture \emph{even in the absence of IIA}. This, in turn, indicates that the IIA is small, which is consistent with previous results obtained by magnetotransport~\cite{Q33,Q44,Q45,Q46}, photoconductivity~\cite{Q47} and the measurements of terahertz transparency~\cite{Q48}. Importantly, the proposed many-particle mechanism describing the observed behavior of LL transitions is intrinsic to HgTe QWs of arbitrary crystallographic orientation, including (110) and (111) QWs, where IIA does not induce anticrossing of the zero-mode LLs.


\begin{acknowledgments}
This work was supported by the Occitanie region through the ``Occitanie Terahertz Platform'' and by the French Agence Nationale pour la Recherche for TEASER project (ANR-24-CE24-4830) and by the CNRS for the Tremplin 2024 STEP project. Finally, we wish to express our gratitude to B. Mongellaz for his invaluable support in managing the helium recovery service, especially at times when reliable assistance was unexpectedly lacking. The experimental data that support the findings of this study are publicly available in Ref.~\cite{Ruffenach2025_data}.
\end{acknowledgments}

%

\newpage
\clearpage
\setcounter{equation}{0}
\setcounter{figure}{0}
\setcounter{table}{0}
\renewcommand{\thefigure}{S\arabic{figure}}

\onecolumngrid
\section*{Supplemental Materials}
\maketitle
\onecolumngrid

\subsection{A. The waterfall plot of magnetoabsorption spectra presented in the main text}
\begin{figure*}[h!]
\includegraphics [width=0.525\columnwidth, keepaspectratio] {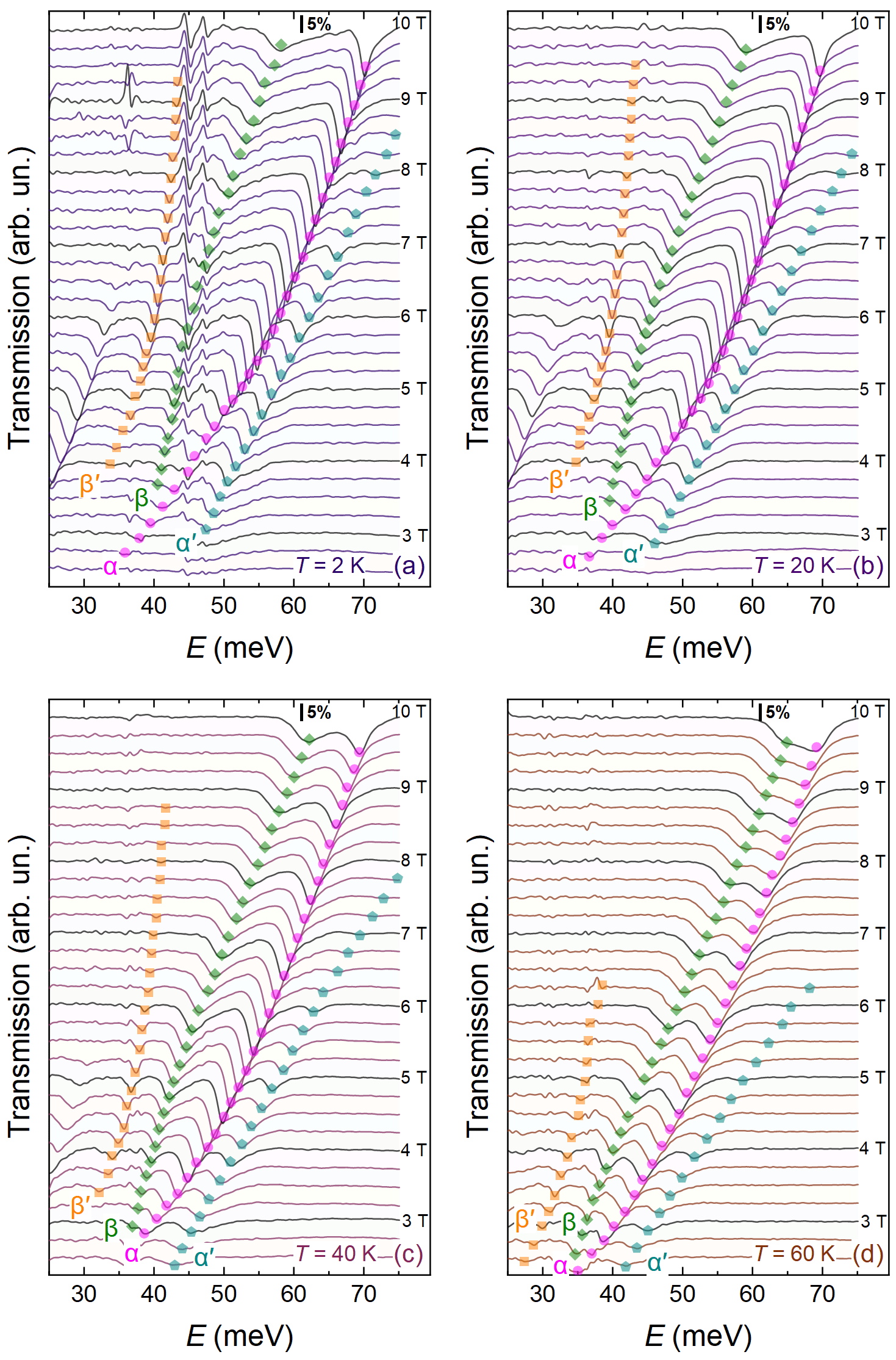} 
\caption{\label{Fig:SM1b} The waterfall plot of magnetoabsorption spectra presented in Fig.~3 in the main text. The resonant energies
for $\alpha$, $\beta$, $\alpha'$ and $\beta'$ transitions are marked by the symbols.}
\end{figure*}

Figure~\ref{Fig:SM1b} shows the magnetoabsorption spectra provided in the main text, plotted in the waterfall format.

\subsection{B. Magnetoabsorption spectra for the temperatures not presented in the main text}
\begin{figure*}[h!]
\includegraphics [width=0.85\columnwidth, keepaspectratio] {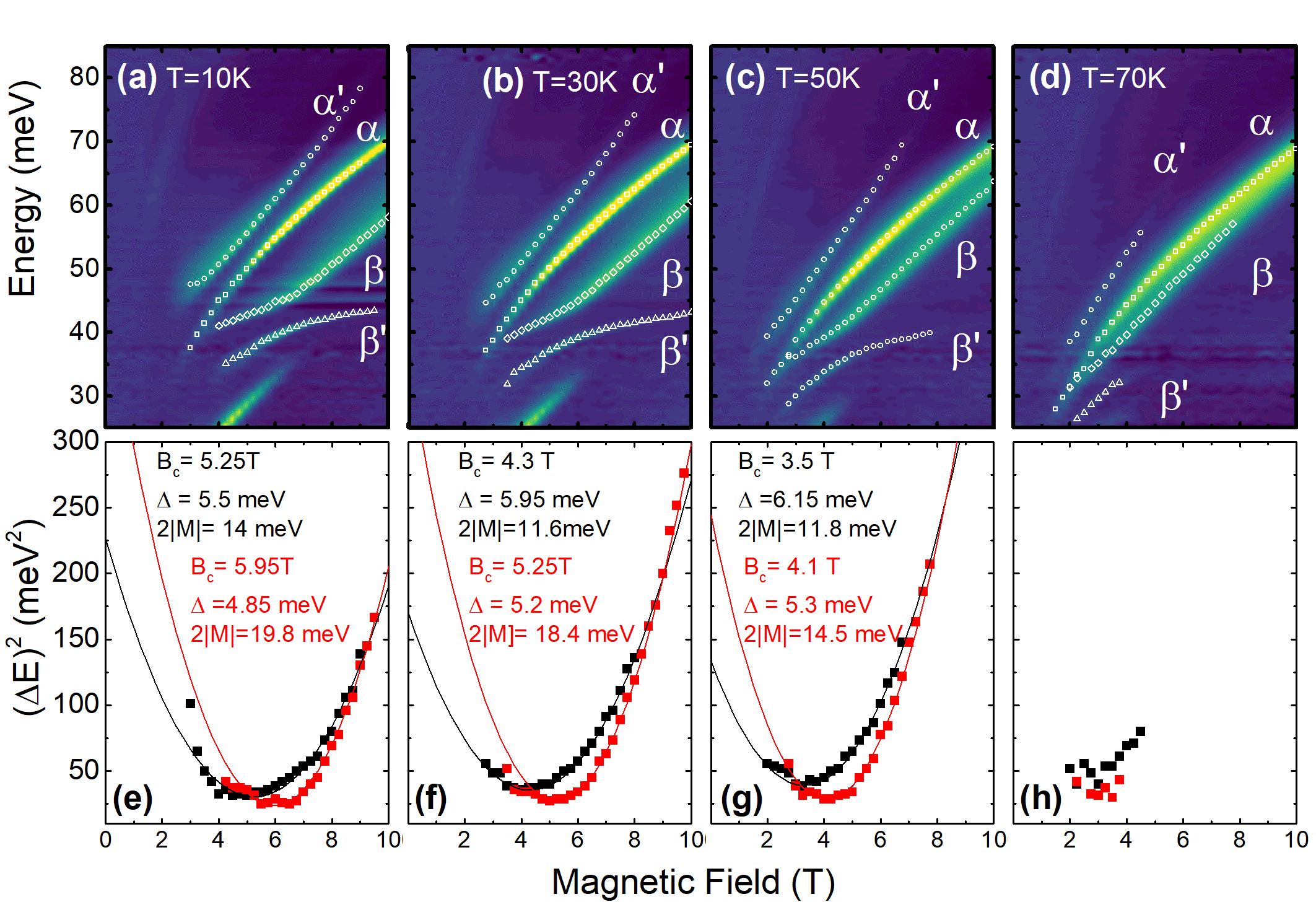} 
\caption{\label{Fig:SM1} \textbf{(a)-(d)} Color maps of magnetoabsorption showing $\alpha$, $\alpha'$, $\beta$, $\beta'$ LL transitions as a function of magnetic field, measured at different temperatures: (a)~$10$~K, (b)~$30$~K, (c)~$50$~K, (d)~$70$~K. The symbols represent position of the magnetoabsorption lines, whose resonant energies are used in the evaluation of the energy difference. \textbf{(e)-(h)} Square of the energy difference for $|\hbar\omega_{\alpha'}-\hbar\omega_{\alpha}|$ and $|\hbar\omega_{\beta'}-\hbar\omega_{\beta}|$ at the same temperatures as in the respective top panels. The black and red solid curves are the fitting to Eq.~(2) in the main text for the pairs ($\alpha$,~$\alpha'$) and ($\beta$,~$\beta'$), respectively. The data at $70$~K does not allow the effective fitting.}
\end{figure*}

Figure~\ref{Fig:SM1} shows the magnetoabsorption spectra measured in the Faraday configuration at several temperatures: (a) 10 K, (b) 30 K, (c) 50 K, (d) 70 K -- these temperatures are not presented in the main text. The bottom panels provide the fitting of the energy difference by Eq.~(2) for the ($\alpha$, $\alpha'$) and ($\beta$, $\beta'$) pairs of the Landau level (LL) transitions. The observation of $\alpha'$ and $\beta'$ LL transitions in a limited range of magnetic fields does not allow the effective fitting of the energy difference $|\hbar\omega_{\alpha'}-\hbar\omega_{\alpha}|$ and $|\hbar\omega_{\beta'}-\hbar\omega_{\beta}|$ at $70$~K.

\subsection{C. Effect of IIA on the anticrossing of zero-mode LLs in HgTe~QWs of arbitrary orientations}
In this section we address analytically the effect of interface inversion asymmetry (IIA) on the anticrossing of zero-mode LLs in HgTe quantum well (QW) of arbitrary orientation. The microscopic origin of the IIA is the anisotropy of chemical bonds at the interface between two bulk materials, that results in mixing of light-hole and heavy-hole Bloch functions~\cite{SM1}. Thus, the minimal multiband Hamiltonian for describing the IIA in HgTe QWs should must include the mutual coupling between the $\Gamma_6$ and $\Gamma_8$ bands of bulk semiconductors~\cite{SMbook1}. Although for quantitative description of the position of electron-like subbands in narrow HgTe QWs it is necessary to take into account the additional ``split-off'' $\Gamma_7$ band (see Supplementary material to Ref.~\cite{SM2}), we will not take it into account in the context of the problem under consideration. We emphasize that taking into account the $\Gamma_7$ contribution only complicates the derivation of the analytical expression for the anticrossing gap without changing it.

In the basis set of Bloch amplitudes in the sequence $|\Gamma_6,+1/2\rangle$,$|\Gamma_6,-1/2\rangle$, $|\Gamma_8,+3/2\rangle$, $|\Gamma_8,+1/2\rangle$, $|\Gamma_8,-1/2\rangle$, $|\Gamma_8,-3/2\rangle$~\cite{SMbook1}, the 6-band \textbf{k$\cdot$p} Hamiltonian can be presented in the form
\begin{equation}
\label{eq:A1}
H_{3D}=\begin{pmatrix}
H_{cc}^{\mathbf{k}\cdot\mathbf{p}} & H_{cv}^{\mathbf{k}\cdot\mathbf{p}} \\{H_{cv}^{\mathbf{k}\cdot\mathbf{p}} }^{\dag} & H_{vv}^{\mathbf{k}\cdot\mathbf{p}}\end{pmatrix}+
H^{\mathrm{IIA}},
\end{equation}
where the first term describes the effets of quantum confinement and strain (due to lattice-mismatch in the QW layers), while the second term represents the IIA contribution~\cite{SM1}. Here, the subscripts ``\emph{cc}'' and ``\emph{vv}'' are labeling the matrices corresponding to the contribution from the $\Gamma_6$ and $\Gamma_8$ bands, respectively, and the matrices with the ``\emph{cv}'' index describe their interband mixing.

Let us start with the first term of $H_{3D}$ describing the quantum confinement and strain effects. Assuming that the QW growth direction coincides with $z$ axis, the block $H_{cc}^{\mathbf{k}\cdot\mathbf{p}}$ in (\ref{eq:A1}) can be written as
\begin{equation}
\label{eq:A2}
H_{cc}^{\mathbf{k}\cdot\mathbf{p}}=\left[E_c(z)+\dfrac{\hbar^{2}\mathbf{k}\left[2F(z)+1\right]\mathbf{k}}{2m_0} + \Xi_{c}\mathrm{Tr}\epsilon\right]I_{2{\times}2},
\end{equation}
where $I_{2{\times}2}$ is the $2{\times}2$ identity matrix, $E_c(z)$ is the conduction band profile, $\mathbf{k}=(k_x,k_y,k_z)$ (note that $k_z=-i\partial/\partial{z}$ as $z$ is the growth direction), $F(z)$ is a parameter accounting for contribution from the remote bands (which are not included in $H_{3D}$), $\Xi_{c}$ is the $\Gamma_6$-band deformation potential constant, and $\epsilon$ is the strain tensor arising due to lattice-mismatch in the QW layers and the sample substrate. The block $H_{cv}^{\mathbf{k}\cdot\mathbf{p}}$ in (\ref{eq:A1}) has the form
\begin{equation}
\label{eq:A3}
H_{cv}^{\mathbf{k}\cdot\mathbf{p}}=\begin{pmatrix}
-\dfrac{\sqrt{2}Pk_{+}}{2} & \dfrac{\sqrt{6}Pk_{z}}{3} & \dfrac{\sqrt{6}Pk_{-}}{6} & 0 \\[6pt]
0 & -\dfrac{\sqrt{6}Pk_{+}}{6} & \dfrac{\sqrt{6}Pk_{z}}{3} & \dfrac{\sqrt{2}Pk_{-}}{2} \end{pmatrix},
\end{equation}
where $P$ is the Kane matrix element, $k_{\pm}=k_x{\pm}ik_y$. The block $H_{vv}^{\mathbf{k}\cdot\mathbf{p}}$ in (\ref{eq:A1}) is given by
\begin{equation}
\label{eq:A4}
H_{vv}^{\mathbf{k}\cdot\mathbf{p}}=E_v(z)I_{4{\times}4}+H^{(i)}_L+H^{(a)}_L+H^{(i)}_{BP}+H^{(a)}_{BP},
\end{equation}
where $I_{4{\times}4}$ is the $4{\times}4$ identity matrix, $E_v(z)$ is the valence band profile, $H^{(i)}_L$, $H^{(a)}_L$, $H^{(i)}_{BP}$ and $H^{(a)}_{BP}$ are the isotropic and anisotropic parts of the Luttinger and Bir-Pikus Hamiltonians~\cite{SMbook1}. The isotropic parts are written as
\begin{eqnarray}
\label{eq:A5}
H^{(i)}_L=\dfrac{\hbar^2}{2m_0}\left[-\mathbf{k}\left(\gamma_1+\dfrac{5}{2}\gamma_2\right)\mathbf{k}+ 2(\mathbf{J}\cdot\mathbf{k})\gamma_2(\mathbf{J}\cdot\mathbf{k})\right],~~\nonumber\\
H^{(i)}_{BP}=\left(a+\dfrac{5}{4}b\right)\mathrm{Tr}\epsilon-b\sum_{\alpha}J_\alpha^2\epsilon_{\alpha\alpha} -d\sum_{\alpha\neq\beta}\{J_\alpha,J_\beta\}_s\epsilon_{\alpha\beta},
\end{eqnarray}
where $\textbf{J}$ is the vector composed of the matrices of the angular momentum $3/2$; $\{J_\alpha,J_\beta\}_s=(J_{\alpha}J_{\beta}+J_{\beta}J_{\alpha})/2$; $a$, $b$, and $d$ are the $\Gamma_8$-band deformation potential constants.

Until now, we have not specified the orientation of Cartesian coordinate system with respect to the main crystallographic axes, because $H_{cc}^{\mathbf{k}\cdot\mathbf{p}}$, $H_{cv}^{\mathbf{k}\cdot\mathbf{p}}$, $H^{(i)}_L$ and $H^{(i)}_{BP}$ retain their forms under rotation of the coordinate system. In what follows, we will deal with terms related to the cubic symmetry of zinc-blende semiconductors, therefore their forms below are valid only if $x\parallel[100]$, $y\parallel[010]$, and $z\parallel[001]$. In this case, the anisotropic terms $H^{(a)}_L$ and $H^{(a)}_{BP}$ in Eq.~(\ref{eq:A4}) are written as:
\begin{eqnarray}
\label{eq:A6}
H^{(a)}_L=\dfrac{\hbar^2}{2m_0}\left[\{J_x,J_y\}_s(\gamma_3-\gamma_2)k_{x}k_{y} +\{J_x,J_z\}_s\{\gamma_3-\gamma_2,k_{z}\}_{s}k_{x}+\{J_y,J_z\}_s\{\gamma_3-\gamma_2,k_{z}\}_sk_{y}\right],\nonumber\\
H^{(a)}_{BP}=-2\left(\dfrac{d}{\sqrt{3}}-b\right)\left[\{J_x,J_y\}_s\epsilon_{xy} +\{J_x,J_z\}_s\epsilon_{xz} + \{J_y,J_z\}_s\epsilon_{yz}\right].~~~~~~~~~~~~~~~~~~~~~
\end{eqnarray}
Importantly, the strain tensor components $\epsilon_{\alpha\beta}$ should be found from the elastic energy minimization that, in turn, also depends on the orientation of the coordinate system with respect to the main cubic axes. Note that all parameters $\gamma_1$, $\gamma_2$,  $a$, $b$, $d$ and $\epsilon_{\alpha\beta}$ in the above expressions are functions of the $z$ coordinate.

Finally, the contribution due to IIA from the two QW interfaces is also anisotropic, which for $x\parallel[100]$, $y\parallel[010]$, and $z\parallel[001]$ can be written as~\cite{SM3}:
\begin{equation*}
H^{\mathrm{IIA}}=\begin{pmatrix}
\hat{0} & \hat{0} \\{ \hat{0} }^{\dag} & H_{vv}^{\mathrm{IIA}}\end{pmatrix},
\end{equation*}
with
\begin{equation}
\label{eq:A9}
H_{vv}^{\mathrm{IIA}}=\mathfrak{I}_{\mathrm{IIA}}
\left[\delta\left(\mathbf{r}\cdot\mathbf{n}+r_i\right)-\delta\left(\mathbf{r}\cdot\mathbf{n}+r_i+d\right)\right]
\left[\{J_x,J_y\}_{s}n_{z} + \{J_y,J_z\}_{s}n_{x} + \{J_z,J_x\}_{s}n_{y}\right],
\end{equation}
where $\mathfrak{I}_{\mathrm{IIA}}$ is constant characterizing the given interface, $\mathbf{n}=(n_x,n_y,n_z)$ is the unit vector perpendicular to the interface. In this case, $\mathbf{r}\cdot\mathbf{n}+r_i=0$ is the equation of the interface plane, and $r_i$ is the distance between the first QW interface and the coordinate origin, while $d$ is the QW width.

To calculate the Landau levels in HgTe/CdHgTe QWs, it is usually sufficient to use only the first term of $H_{3D}$ in Eq.~(\ref{eq:A1}) within in the so-called \emph{axial approximation}, which, along with the isotropic terms $H_{cc}^{\mathbf{k}\cdot\mathbf{p}}$ and $H_{cv}^{\mathbf{k}\cdot\mathbf{p}}$, also takes into account some anisotropic terms of $H_{vv}^{\mathbf{k}\cdot\mathbf{p}}$ that preserve axial symmetry in the QW plane. An explicit form of such an axially symmetric Hamiltonian depends on the QW growth plane~\cite{SM2}.

The main advantage of the axial approximation is that allows obtaining a clear picture of LLs in the magnetic field perpendicular to the QW plane. Particularly, choosing the magnetic vector potential in Landau gauge $\textbf{A}$= (0, $Bx$,0) and using a Peierls substitution
\begin{equation}
\label{eq:A10}
k_{x}=-i\frac{\partial}{\partial x}+\frac{e}{\hbar c}A_{x},~~~~~~~~~~~~k_{y}=-i\frac{\partial}{\partial y}+\frac{e}{\hbar c}A_{y},
\end{equation}
one can introduce the ladder operators $b^{+}$ and $b$:
\begin{equation}
\label{eq:A11}
b^{+}=\frac{a_{B}}{\sqrt{2}}k_{+},~~~~~~~~~~~~
b=\frac{a_{B}}{\sqrt{2}}k_{-},
\end{equation}
where $a_{B}$ is the magnetic length ($a_{B}^{2} =c\hbar/eB$), $e>0$ is the elementary charge, and write the total wave function in the form
\begin{equation}
\label{eq:A12}
\Psi_{n_{z},N,i,{k}}^{(\mathrm{axial})}(x,y,z)=
\begin{pmatrix}
c_{1}^{(\mathrm{axial})}(z,n_{z},N,i)|N,{k}\rangle \\[3pt]
c_{2}^{(\mathrm{axial})}(z,n_{z},N,i)|N+1,{k}\rangle \\[3pt]
c_{3}^{(\mathrm{axial})}(z,n_{z},N,i)|N-1,{k}\rangle \\[3pt]
c_{4}^{(\mathrm{axial})}(z,n_{z},N,i)|N,{k}\rangle \\[3pt]
c_{5}^{(\mathrm{axial})}(z,n_{z},N,i)|N+1,{k}\rangle \\[2pt]
c_{6}^{(\mathrm{axial})}(z,n_{z},N,i)|N+2,{k}\rangle
\end{pmatrix}.
\end{equation}
Here, $n_z$ is the electronic subband index, $N$ is the LL index, $i$ is the ``spin'' index (which labels LLs with the same $N$ -- see Fig.~1 in the main text), while $c_{i}^{(\mathrm{axial})}(z,n_{z},N,i)$ is the envelope of the Bloch function of the corresponding band in the QW growth direction and
\begin{eqnarray}
\label{eq:A13}
|N,{k}\rangle=
\begin{cases}
0,~N<0,\\[3pt]
\dfrac{\exp\left(i{k}y\right)}{\sqrt{2^{N}N!\sqrt{\pi}a_{B}L_{y}}}H_{N}\left(\dfrac{\tilde{x}}{a_{B}}\right)\exp\left(-\dfrac{\tilde{x}^{2}}{2a_{B}^{2}}\right),~N\geq0,
\end{cases}\nonumber\\
\tilde{x}=x-{k}a_{B}^2,~~~~~~~~~~~~~~~~~~~~~~~~~~~~~
\end{eqnarray}
where $L_{y}$ is the sample size along the $y$ axis, $H_{N}$ are the Hermitian polynomials with number $N$, and ${k}$ is the wave-vector projection onto the $y$ axis in the Landau gauge.

The two LLs, identified as \emph{zero-mode LLs} within effective 2D Bernevig-Hughes-Zhang (BHZ) model~\cite{SM4}, are characterized by $N=0$ and $N=-2$ LL indices (see Fig.~1 in the main text). The latter has only a sixth non-zero component (for others, LL index takes the ``negative'' values), representing the contribution of $|\Gamma_8,-3/2\rangle$ Bloch function. Importantly, the non-axial terms of $H_{vv}^{\mathbf{k}\cdot\mathbf{p}}$ in Eq.~(\ref{eq:A4}) have almost no effect on the energy dependence of zero-mode LLs on the magnetic field~\cite{SM5}. This allows to use the wave functions $\left|\Psi_{n_{z},N,{k}}^{(\mathrm{axial})}\right\rangle$ for the anticrossing gap calculations instead of more complex wave functions of the non-axial approximation.

Since other LLs lie far from the possible anticrossing region of zero-mode LLs, for the anticrossing gap calculation, one has just to project $H^{\mathrm{IIA}}$ onto two following states:
\begin{equation}
\label{eq:A14}
\left|\Psi_{E1,0,\tilde{k}}^{(\mathrm{axial})}\right\rangle=
\begin{pmatrix}
c_{1}^{(\mathrm{axial})}(z,E1,0)|0,{k}\rangle \\[3pt]
c_{2}^{(\mathrm{axial})}(z,E1,0)|1,{k}\rangle \\[3pt]
0 \\[3pt]
c_{4}^{(\mathrm{axial})}(z,E1,0)|0,{k}\rangle \\[3pt]
c_{5}^{(\mathrm{axial})}(z,E1,0)|1,{k}\rangle \\[2pt]
c_{6}^{(\mathrm{axial})}(z,E1,0)|2,{k}\rangle
\end{pmatrix},~~~~~~~~~~~~~
\left|\Psi_{H1,-2,{k}}^{(\mathrm{axial})}\right\rangle=
\begin{pmatrix}
0 \\[3pt]
0 \\[3pt]
0 \\[3pt]
0 \\[3pt]
0 \\[2pt]
c_{6}^{(\mathrm{axial})}(z,H1,-2)|0,{k}\rangle
\end{pmatrix}.
\end{equation}

\begin{figure*}
\includegraphics [width=0.5\columnwidth, keepaspectratio] {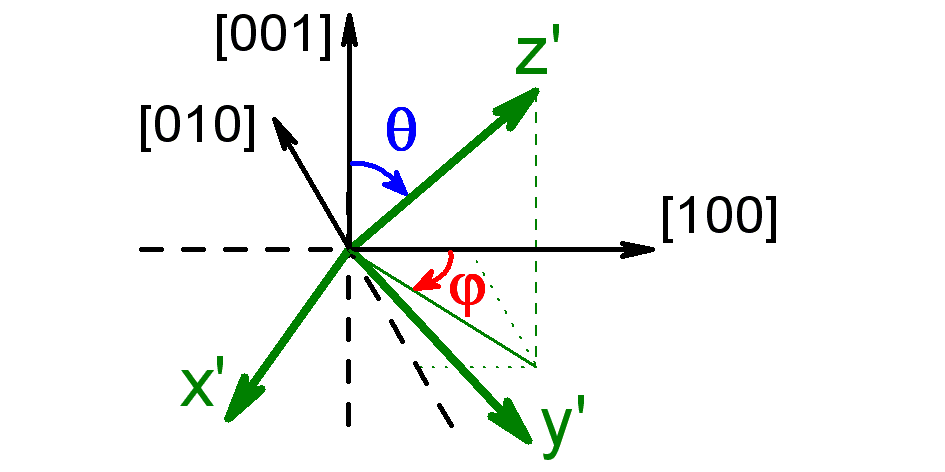} 
\caption{\label{Fig:SM2} Orientation of the new Cartesian coordinate system with respect to the main crystallographic axes. The QW growth direction concide with the $z'$ axis, whose orientation is defined by angles $\varphi$ and $\theta$ with respect to the main cubic axes. Note that the positive values of $\varphi$ and $\theta$ correspond to clockwise rotations.}
\end{figure*}

After the projection, we get an effective 2$\times$2 Hamiltonian, whose off-diagonal terms will describe the anticrossing gap. In order to derive the form of this effective 2$\times$2 Hamiltonian for the QW of arbitrary orientation, we rewrite $H^{\mathrm{IIA}}$ in another Cartesian coordinate system, where the new $z'$ axis no longer coincides with [001] direction (see Fig.~\ref{Fig:SM2}). Simultaneously with the transition from $\mathbf{k}$ to $\mathbf{k'}$ (and from $\mathbf{r}$ to $\mathbf{r'}$ by means of the same transformation)
\begin{equation}
\label{eq:A15}
\begin{pmatrix}
k_x \\
k_y \\
k_z
\end{pmatrix}=
\begin{pmatrix}
\cos\varphi & -\sin\varphi & 0 \\
\cos\theta\sin\varphi & \cos\theta\cos\varphi & -\sin\theta \\
\sin\theta\sin\varphi & \sin\theta\cos\varphi & \cos\theta
\end{pmatrix}
\begin{pmatrix}
k_{x'} \\
k_{y'} \\
k_{z'}
\end{pmatrix}
\end{equation}
one should also apply a unitary transformation~\cite{SM6}:
\begin{equation}
\label{eq:A16}
{H^{\mathrm{IIA}}}'=\hat{\Gamma}^{-1}H^{\mathrm{IIA}}\hat{\Gamma},
\end{equation}
where
\begin{equation}
\label{eq:A17}
\hat{\Gamma}=
\begin{pmatrix}
\exp\left[-i\dfrac{\sigma_{z}}{2}\varphi\right]\exp\left[-i\dfrac{\sigma_{x}}{2}\theta\right] &  0 \\
0 & \exp\left[-iJ_{z}\varphi\right]\exp\left[-iJ_{x}\theta\right]
\end{pmatrix}.
\end{equation}
Here and further, $\sigma_{x}$, $\sigma_{y}$ and $\sigma_{z}$ are Pauli matrices.

After routine mathematics, the IIA-induced term $H_{vv}^{\mathrm{IIA}}$ in Eq.~(\ref{eq:A9}) can be written in the form:
\begin{multline}
\label{eq:A18}
H_{vv}^{\mathrm{IIA}}=\mathfrak{I}_{\mathrm{IIA}}
\left[\delta\left(z+r_i\right)-\delta\left(z+r_i+d\right)\right]~\\
\times\bigg[\cos{2\theta}\cos{2\varphi}\{J_x,J_y\}_{s}+\sin{\theta}\left(2-3\sin^2{\theta}\right)\sin{2\varphi}\{J_y,J_z\}_{s}
-\sin{2\theta}\cos{2\varphi}\{J_z,J_x\}_{s}~\\
+\cos{\theta}\sin{2\varphi}\dfrac{J_x^2-J_y^2}{2}+\dfrac{3}{2}\sin{\theta}\sin{2\theta}\sin{2\varphi}\dfrac{J_y^2-J_z^2}{2}\bigg],
\end{multline}
where strokes are omitted for simplicity. As seen from the form of the wave functions~(\ref{eq:A14}) of the zero-mode LLs, $H_{vv}^{\mathrm{IIA}}$ opens the anticrossing gap, described by the operator
\begin{equation}
\label{eq:A19}
\hat{\Delta}_{\mathrm{a-c}}^{\mathrm{IIA}}=\dfrac{1}{2}\begin{pmatrix}
\Delta^{\mathrm{IIA}} & 0 \\
0 & {\Delta^{\mathrm{IIA}}} ^{\dag}
\end{pmatrix}
\left[\sigma_{x}\dfrac{\cos{\theta}\left(3\cos^2{\theta}-1\right)\sin{2\varphi}}{2}+\sigma_{y}\cos{2\theta}\cos{2\varphi}\right],
\end{equation}
where
\begin{equation*}
\Delta^{\mathrm{IIA}}=\dfrac{\sqrt{3}}{3}\mathfrak{I}_{\mathrm{IIA}}\left[\left(c_{4}^{(\mathrm{axial})}(-r_i,E1,0)\right)^{*}c_{6}^{(\mathrm{axial})}(-r_i,H1,-2)-
\left(c_{4}^{(\mathrm{axial})}(-r_i-d,E1,0)\right)^{*}c_{6}^{(\mathrm{axial})}(-r_i-d,H1,-2)\right].
\end{equation*}
Here, the asterisk and ``$\dag$'' denote the complex and  Hermitian conjugation, respectively.

Taking into account Eqs.~(\ref{eq:A19}), the projected Hamiltonian onto the basis functions $\left|\Psi_{E1,0,\tilde{k}}^{(\mathrm{axial})}\right\rangle$ and $\left|\Psi_{H1,-2,\tilde{k}}^{(\mathrm{axial})}\right\rangle$ of zero-mode LLs in  Eq.~(\ref{eq:A14}) is written as
\begin{equation}
\label{eq:A21}
\hat{H}_{\mathrm{zmode}}=\begin{pmatrix}
E_{E1,0} & 0 \\
0 & E_{H1,-2}
\end{pmatrix}
+\hat{\Delta}_{\mathrm{a-c}}^{\mathrm{IIA}}+
\begin{pmatrix}
\Lambda_{E1}^{\mathrm{IIA}} & 0 \\
0 & \Lambda_{H1}^{\mathrm{IIA}}
\end{pmatrix}{G}(\varphi,\theta),
\end{equation}
where
\begin{equation}
\label{eq:A21a0}
{G}(\varphi,\theta)=\sin{2\theta}\sin{\theta}\sin{2\varphi}.
\end{equation}
In Eq.~(\ref{eq:A21}), $E_{E1,0}$ and $E_{H1,-2}$ are the energies of ``unperturbed'' zero-mode LLs within the axial approximation, while $\Lambda_{E1}^{\mathrm{IIA}}$ and $\Lambda_{H1}^{\mathrm{IIA}}$ are \emph{real} constants, describing the diagonal corrections to the energies of zero-mode LLs:
\begin{multline}
\label{eq:A21a1}
\Lambda_{E1}^{\mathrm{IIA}}=\dfrac{3}{8}\mathfrak{I}_{\mathrm{IIA}}\biggl[\left|c_{4}^{(\mathrm{axial})}(-r_i,E1,0)\right|^2+
\left|c_{5}^{(\mathrm{axial})}(-r_i,E1,0)\right|^2-\left|c_{6}^{(\mathrm{axial})}(-r_i,E1,0)\right|^2~\\
-\left|c_{4}^{(\mathrm{axial})}(-r_i-d,E1,0)\right|^2-\left|c_{5}^{(\mathrm{axial})}(-r_i-d,E1,0)\right|^2
+\left|c_{6}^{(\mathrm{axial})}(-r_i-d,E1,0)\right|^2\biggr],
\end{multline}
\begin{equation}
\label{eq:A21a2}
\Lambda_{H1}^{\mathrm{IIA}}=\dfrac{3}{8}\mathfrak{I}_{\mathrm{IIA}}\left[-\left|c_{6}^{(\mathrm{axial})}(-r_i,H1,-2)\right|^2
+\left|c_{6}^{(\mathrm{axial})}(-r_i-d,H1,-2)\right|^2\right].
\end{equation}

The eigenvalues of $\hat{H}_{\mathrm{zmode}}$ represent the energies of zero-mode LLs, ``modified'' by the IIA:
\begin{equation}
\label{eq:A22}
\epsilon^{(\pm)}_0=\dfrac{{E}_{E1,0}+{E}_{H1,-2}}{2}+\dfrac{\Lambda_{E1}^{\mathrm{IIA}}+\Lambda_{H1}^{\mathrm{IIA}}}{2}{G}(\varphi,\theta)\pm\sqrt{ \left(\dfrac{E_{E1,0}-E_{H1,-2}}{2}+\dfrac{\Lambda_{E1}^{\mathrm{IIA}}-\Lambda_{H1}^{\mathrm{IIA}}}{2}{G}(\varphi,\theta)\right)^2+\dfrac{|\Delta^{\mathrm{IIA}}|^2}{4}{F}(\varphi,\theta)^2},
\end{equation}
where $\Lambda_{E1}^{\mathrm{IIA}}$, $\Lambda_{H1}^{\mathrm{IIA}}$ and $\Delta^{\mathrm{IIA}}$ have very weak dependence on the QW growth orientation, while $F(\varphi,\theta)$ has the following form:
\begin{equation}
\label{eq:A23}
F(\varphi,\theta)=\sqrt{\cos^2{2\theta}\cos^2{2\varphi}+\dfrac{\cos^2{\theta}\left(3\cos^2{\theta}-1\right)^2\sin^2{2\varphi}}{4}}.
\end{equation}
Assuming a linear magnetic-field-dependence for the difference $E_{E1,0}-E_{H1,-2}$
\begin{equation}
\label{eq:A24}
E_{E1,0}-E_{H1,-2}{\simeq}2M\left(1-\dfrac{B}{B_c}\right),
\end{equation}
where $M$ is a mass parameter, whose sign defines the band inversion between \emph{E}1 and \emph{H}1 subband; and $B_c$ is critical magnetic field, corresponding to the crossing of zero-mode LLs in the absence of IIA, we get
\begin{equation}
\label{eq:A25}
\left|\epsilon^{(+)}_0-\epsilon^{(-)}_0\right|=\sqrt{\left(2\widetilde{M}\right)^2\left(1-\dfrac{B}{\widetilde{B}_c}\right)^2+\Delta^2},
\end{equation}
where
\begin{equation}
\label{eq:A26}
\widetilde{M}=M\left(1+\dfrac{\Lambda_{E1}^{\mathrm{IIA}}-\Lambda_{H1}^{\mathrm{IIA}}}{2M}\right){G}(\varphi,\theta),
\end{equation}
\begin{equation}
\label{eq:A27}
\widetilde{B}_c={B}_c\left(1+\dfrac{\Lambda_{E1}^{\mathrm{IIA}}-\Lambda_{H1}^{\mathrm{IIA}}}{2M}\right){G}(\varphi,\theta),
\end{equation}
\begin{equation}
\label{eq:A27}
\Delta=|\Delta^{\mathrm{IIA}}|{F}(\varphi,\theta).
\end{equation}
As clear, at $\theta=0$ for (001) HgTe QWs, ${G}(\varphi,\theta)=0$, ${F}(\varphi,\theta)=1$, and Eq.~(\ref{eq:A25}) takes the form of Eq.~(2) in the main text.

Importantly, for the QWs grown along [011] ($\varphi=0$, $\theta=\pi/4$) and [111] ($\varphi=-\pi/4$, $\theta=\arccos(1/\sqrt{3})$) directions, $F(\varphi,\theta)$ vanishes. Thus, in the QWs of these orientations (as well as in equivalent ones), the IIA does not induce the anticrossing of zero-mode LLs. The same conclusion was previously made for the anticrossing of zero-mode LLs induced by the bulk inversion asymmetry (BIA)~\cite{SM6}.

\subsection{D. Many-particle hybridization of LL transitions}
This section presents a theoretical justification of the many-particle mechanism that explains the observed behavior of optical transitions from zero-mode LLs by the influence of electron-electron (\emph{e-e}) interaction on the hybridization of LL transitions. For simplicity, we will not take into account the possible influence of IIA, and as in the previous section, we will treat single-particle LLs in HgTe QW within the \emph{axial approximation}. Taking into account all of the above, the many-particle Hamiltonian of HgTe QW in the second quantized representation can be written in the form:
\begin{eqnarray}
\label{eq:C1}
\hat{\mathcal{H}}=\int\limits_{-\infty}^{+\infty} dz \int d^2\vec{\rho}\hat{\Psi}^{+}(\vec{\rho},z)H_{3D}\hat{\Psi}(\vec{\rho},z)+\hat{H}_{int},\notag~~~~~~~~~~~~~~~~~~~~~~~~~~~~~~~~~~~\\
\hat{H}_{int}=\dfrac{1}{2}\int\limits_{-\infty}^{+\infty} dz_1 \int\limits_{-\infty}^{+\infty} dz_2 \int d^2\vec{\rho}_1\int d^2\vec{\rho}_2\hat{\Psi}^{+}(\vec{\rho}_1,z_1)\hat{\Psi}^{+}(\vec{\rho}_2,z_2)
V(\left|\vec{\rho}_1-\vec{\rho}_2\right|,z_1,z_2)\hat{\Psi}(\vec{\rho}_2,z_2)\hat{\Psi}(\vec{\rho}_1,z_1),~
\end{eqnarray}
where $\vec{\rho}=(x,y)$ is the vector lying in the QW plane, and $V(\left|\vec{\rho_1}-\vec{\rho_2}\right|,z_1,z_2)$ is the Coulomb Green function in a three-layer medium, describing the interaction between the charges at points $(\vec{\rho_1},z_1)$ and $(\vec{\rho_2},z_2)$~\cite{SM7,SM8}.

In Eq.~(\ref{eq:C1}), we have introduced the field operators $\hat{\Psi}(\vec{\rho},z)$ and $\hat{\Psi}^{+}(\vec{\rho},z)$ defined by the fermion creation and annihilation operators $a_{n,k,i}$ and $a^{+}_{n,k,i}$, respectively, and by the single-electron wave functions of Hamiltonian $H_{3D}$:
\begin{eqnarray}
\label{eq:C2}
\hat{\Psi}(\vec{\rho},z)=\sum_{n,k,i}\Psi_{n,i,{k}}^{(\mathrm{axial})}(\vec{\rho},z)
a_{n,k,i},~~~~\notag\\
\hat{\Psi}^{+}(\vec{\rho},z)=\sum_{n,k,i}\left(\Psi_{n,i,{k}}^{(\mathrm{axial})}(\vec{\rho},z)\right)^{+}a^{+}_{n,k,i},
\end{eqnarray}
where the upper sign ``+'' denotes the Hermitian conjugation, while multi-index $n=\left(n_z,N\right)$ is introduced for brevity.

As mentioned in the main text, each of the inter-LL excitations in the quantum Hall regime with well-separated LLs can be considered as neutral collective modes described in terms of \emph{magnetic excitons} (MEs)~\cite{SM9,SM10,SM11v0,SM11,SM12} -- i.e., a bound state of a hole in a filled LL and an electron at an empty level. This type of excitation can be naturally described by introducing the magnetic-exciton creation operator~\cite{SM11v0,SM11,SM12}:
\begin{equation}
\label{eq:C3}
A_{n,n',i,i'}^{+}(\vec{k})=\sum_{p}e^{ik_x(p+k_y/2)a^2_B}a^{+}_{n,p,i}a_{n',p+k_y,i'}.
\end{equation}
that satisfies the following commutation relation:
\begin{multline}
\label{eq:C4}
\left[A_{n_1,n_2,i_1,i_2}^{+}(\vec{k}_1),A_{n_3,n_4,i_3,i_4}^{+}(\vec{k}_2)\right]=e^{-\frac{i}{2}a^2_{B}{[\vec{k}_1\times\vec{k}_2]}_z}A_{n_1,n_4,i_1,i_4}^{+}(\vec{k}_1+\vec{k}_2)\delta_{n_2,n_3}\delta_{i_2,i_3}\\
-\delta_{n_1,n_4}\delta_{i_1,i_4}
e^{\frac{i}{2}a^2_{B}{[\vec{k}_1\times\vec{k}_2]}_z}A_{n_3,n_2,i_3,i_2}^{+}(\vec{k}_1+\vec{k}_2),
\end{multline}
where $a_{B}$ is the magnetic length ($a_{B}^{2}=c\hbar/eB$).

By mathematical calculations, it can be directly shown that $\hat{H}_{int}$ in Eq.~(\ref{eq:C1}) can be represented in terms of ME operators as follows:
\begin{multline}
\label{eq:C5}
\hat{H}_{int}=\dfrac{1}{2}\sum_{\substack{n_{1}...n_{4}\\i_{1}...i_{4}}}\int \frac{d^2\vec{q}}{(2\pi)^2}\tilde{V}^{(i_{1},i_{2},i_{3},i_{4})}_{n_{1},n_{2},n_{3},n_{4}}(\vec{q})
A_{n_1,n_4,i_1,i_4}^{+}(\vec{q})A_{n_2,n_3,i_2,i_3}^{+}(-\vec{q})-\\
-\dfrac{1}{2}\sum_{\substack{n_{1},n_{2},n_{3}\\i_{1},i_{2},i_{3}}}\int \frac{d^2\vec{q}}{(2\pi)^2}\tilde{V}^{(i_{1},i_{2},i_{2},i_{3})}_{n_{1},n_{2},n_{2},n_{3}}(\vec{q})A_{n_1,n_3,i_1,i_3}^{+}(0),
\end{multline}
where the matrix element $\tilde{V}^{(i_{1},i_{2},i_{3},i_{4})}_{n_{1},n_{2},n_{2},n_{3}}(\vec{q})$ is defined as
\begin{equation}
\label{eq:C6}
\tilde{V}^{(i_{1},i_{2},i_{3},i_{4})}_{n_{1},n_{2},n_{3},n_{4}}(\vec{q})=\int\limits_{-\infty}^{+\infty} dz_1 \int\limits_{-\infty}^{+\infty} dz_2 \tilde{D}(q,z_1,z_2)e^{-q^{2}a_{B}^{2}/2}
\tilde{G}^{(i_{1},i_{4})}_{n_{1},n_{4}}(\vec{q},z_{1},z_{1})\tilde{G}^{(i_{2},i_{3})}_{n_{2},n_{3}}(-\vec{q},z_{2},z_{2})
\end{equation}
with $\tilde{D}(q,z_1,z_2)$ being the Fourier transform for the Coulomb Green function:
\begin{equation}
\label{eq:C7}
V(|\vec{r}_1-\vec{r}_2|,z_1,z_2)=\int\frac{d^{2}\vec{q}}{(2\pi)^{2}}\tilde{D}(q,z_1,z_2)e^{i\vec{q}(\vec{r}_1-\vec{r}_2)},
\end{equation}
and $\tilde{G}^{(i_{1},i_{2})}_{n_{1},n_{2}}(\vec{q},z,z)$ being written in the form:
\begin{equation}
\label{eq:C8}
\tilde{G}^{(i_1,i_2)}_{n_1,n_2}(\vec{q},z,z)= \tilde{L}^{(i_1,i_2)}_{n_1,n_2}\left(\frac{q^{2}a_{B}^{2}}{2},z,z\right)
\begin{cases}
\left[\dfrac{(iq_x+q_y)a_B}{\sqrt{2}}\right]^{n_1-n_2},~n_1\geq n_2,\\[3pt]
\left[\dfrac{(iq_x-q_y)a_B}{\sqrt{2}}\right]^{n_2-n_1},~n_1<n_2,
\end{cases}
\end{equation}
where $q=\sqrt{q_{x}^{2}+q_{y}^{2}}$. Additionally, $\tilde{L}_{n,n'}^{(i,i')}\left(x,z,z\right)$ in Eq.~\eqref{eq:C8} is determined by the single-particle wave-function $\Psi_{n,i,{k}}^{(\mathrm{axial})}(\vec{\rho},z)$ in Eq.~(\ref{eq:A12}):
\begin{multline}
\label{eq:C9}
\tilde{L}_{n,n'}^{(i,i')}\left(x,z,z\right)=
\left({c_{1}^{(\mathrm{axial})}(z,n,i)}^{*}c_{1}^{(\mathrm{axial})}(z,n',i')+{c_{4}^{(\mathrm{axial})}(z,n,i)}^{*}c_{4}^{(\mathrm{axial})}(z,n',i')
\right)\sqrt{\dfrac{\tilde{n}_{1}!}{\tilde{n}_{2}!}}L_{\tilde{n}_{1}}^{\tilde{n}_{2}-\tilde{n}_{1}}\left(x\right)+
\\
+\left({c_{2}^{(\mathrm{axial})}(z,n,i)}^{*}c_{2}^{(\mathrm{axial})}(z,n',i')+{c_{5}^{(\mathrm{axial})}(z,n,i)}^{*}c_{5}^{(\mathrm{axial})}(z,n',i') \right)\sqrt{\dfrac{(\tilde{n}_{1}+1)!}{(\tilde{n}_{2}+1)!}}L_{\tilde{n}_{1}+1}^{\tilde{n}_{2}-\tilde{n}_{1}}\left(x\right)+
\\
+{c_{3}^{(\mathrm{axial})}(z,n,i)}^{*}c_{3}^{(\mathrm{axial})}(z,n',i') \sqrt{\dfrac{(\tilde{n}_{1}-1)!}{(\tilde{n}_{2}-1)!}}L_{\tilde{n}_{1}-1}^{\tilde{n}_{2}-\tilde{n}_{1}}\left(x\right)
+{c_{6}^{(\mathrm{axial})}(z,n,i)}^{*}c_{6}^{(\mathrm{axial})}(z,n',i')
\sqrt{\dfrac{(\tilde{n}_{1}+2)!}{(\tilde{n}_{2}+2)!}}L_{\tilde{n}_{1}+2}^{\tilde{n}_{2}-\tilde{n}_{1}}\left(x\right),
\end{multline}
where $L_{\tilde{n}_{1}}^{\tilde{n}_{2}-\tilde{n}_{1}}\left(x\right)$ are the associated Laguerre polynomials, $\tilde{n}_{1}=$min$(N,N')$ and $\tilde{n}_{2}=$max$(N,N')$. We recall that here $n$ is a multi-index represented as $n=\left(n_z,N\right)$.

Taking into account Eqs.~(\ref{eq:C5})--(\ref{eq:C9}), the ME energy $\hbar\omega_{n,n'}^{(i,i')}(\vec{k})$ with respect to the energy of the ground state $|0\rangle$ can be found from the following equation~\cite{SM13}:
\begin{equation}
\label{eq:C10}
\hbar\omega_{n,n'}^{(i,i')}(\vec{k})A_{n,n',i,i'}^{+}(\vec{k})|0\rangle=\left(E_{n}^{(i)}-E_{n'}^{(i')}\right)A_{n,n',i,i'}^{+}(\vec{k})|0\rangle
+\left[\hat{H}_{int},A_{n,n',i,i'}^{+}(\vec{k})\right]|0\rangle.
\end{equation}
Now using the commutation relations of Eq.~(\ref{eq:C4}) for the ME operators together with the standard rules of Hartree-Fock approximation
\begin{eqnarray}
\label{eq:C11}
\langle0|a^{+}_{n_1,p_1,i_1}a_{n_2,p_2,i_2}|0\rangle=\delta_{n_1,n_2}\delta_{p_1,p_2}\delta_{i_1,i_2},\notag\\
\sum_{p}\langle0|a^{+}_{n,p,i}a_{n,p,i}|0\rangle=\nu_{n}^{(i)},~~~~~~~~
\end{eqnarray}
where $\nu_{n}^{(i)}$ is the filling factor for LL ($n$, $i$), after rather tedious mathematical transformations,
we obtain the following expression for the commutator in the right-hand part of Eq.~(\ref{eq:C10}):
\begin{multline}
\label{eq:C12}
\left[\hat{H}_{int},A_{n,n',i,i'}^{+}(\vec{k})\right]|0\rangle=
\sum_{n_{2},i_{2}}\nu_{n_2}^{(i_2)}\left(\dfrac{\tilde{V}^{(i,i_2,i_{2},i)}_{n,n_2,n_2,n}(0)}{2\pi}-\dfrac{\tilde{V}^{(i',i_2,i_{2},i')}_{n',n_2,n_2,n'}(0)}{2\pi}\right)A_{n,n',i,i'}^{+}(\vec{k})|0\rangle-\\
-\sum_{n_{2},i_{2}}\nu_{n_2}^{(i_2)}\left(\tilde{E}^{(i,i_2,i,i_{2})}_{n,n_2,n,n_2}(0)-\tilde{E}^{(i',i_2,i',i_{2})}_{n',n_2,n',n_2}(0)\right)A_{n,n',i,i'}^{+}(\vec{k})|0\rangle
-(\nu_{n}^{(i)}-\nu_{n'}^{(i')})\sum_{\substack{n_1,n_4\\i_1,i_4}}\dfrac{\tilde{V}^{(i_1,i',i,i_4)}_{n_1,n',n,n_4}(\vec{k})}{2\pi}A_{n_1,n_4,i_1,i_4}^{+}(\vec{k})|0\rangle+\\
+(\nu_{n}^{(i)}-\nu_{n'}^{(i')})\sum_{\substack{n_1,n_2\\i_1,i_2}}\tilde{E}^{(i',i_1,i,i_2)}_{n',n_1,n,n_2}(\vec{k})A_{n_1,n_2,i_1,i_2}^{+}(\vec{k})|0\rangle
\end{multline}
where
\begin{equation}
\label{eq:C13}
\tilde{E}^{(i_1,i_2,i_3,i_{4})}_{n_1,n_2,n_3,n_4}(\vec{k})=\int \dfrac{d^2\vec{q}}{(2\pi)^2}\tilde{V}^{(i_1,i_2,i_3,i_{4})}_{n_1,n_2,n_3,n_4}(\vec{q})e^{i a^2_{B}{[\vec{q}\times\vec{k}]}_z}.
\end{equation}

From Eq.~(\ref{eq:C12}) it is clear that the electron-electron interaction mixes infinite number of magnetic excitons, therefore Eq.~(\ref{eq:C10}) can be solved only approximately in a limited basis of the exciton states. Since the energies of transition pairs ($\alpha$, $\alpha'$) and ($\beta$, $\beta'$) transitions are fairly well separated from each other and from other LL transitions in the vicinity of $B_c$ (see Fig.~2 in the main text), to describe their hybridization we can limit ourselves to considering only the basic ME states within each pair. Namely,
\begin{eqnarray}
\label{eq:basisA}
|\alpha\rangle=A_{1,0,a,a}^{+}(\vec{k})\,|0\rangle,~~~~\notag\\
|\alpha'\rangle=A_{1,-2,a,a}^{+}(\vec{k})\,|0\rangle,~~
\end{eqnarray}
for $\alpha$ and $\alpha'$ LL transitions, and
\begin{eqnarray}
\label{eq:basisB}
|\beta\rangle=A_{-1,-2,a,a}^{+}(\vec{k})\,|0\rangle,\notag\\
|\beta'\rangle=A_{-1,0,a,a}^{+}(\vec{k})\,|0\rangle,~
\end{eqnarray}
for $\beta$ and $\beta'$ LL transitions, where the index ``$a$'' denotes the LL with the lowest energy for the given $N$ and $n_z$ (see Fig.~1 in the main text).

As a result, the effective Hamiltonian describing the hybridization of ME basic states in the long-wave limit, whose eigenvalues determine the resonant energies of $\alpha$ and $\alpha'$ LL transitions, can be represented as:
\begin{equation}
\label{eq:C14}
\hat{H}_{\mathrm{ME}\alpha\alpha'}=
\begin{pmatrix}
\hbar\omega_{\alpha}^{(0)}+\delta_{\alpha\alpha}^{(e-e)} & \left\{\nu_{0}^{(a)}-\nu_{1}^{(a)}\right\}\Delta_{\alpha\alpha'}^{(e-e)}/2 \\
\left\{\nu_{-2}^{(a)}-\nu_{1}^{(a)}\right\}\Delta_{\alpha\alpha'}^{(e-e)}/2 & \hbar\omega_{\alpha'}^{(0)}+\delta_{\alpha'\alpha'}^{(e-e)}
\end{pmatrix},
\end{equation}
where $\hbar\omega_{\alpha}^{(0)}=E_{1}^{(a)}-E_{0}^{(a)}$ and $\hbar\omega_{\alpha'}^{(0)}=E_{1}^{(a)}-E_{-2}^{(a)}$, while
\begin{equation}
\label{eq:C15}
\dfrac{\Delta_{\alpha\alpha'}^{(e-e)}}{2}=\tilde{E}^{(a,a,a,a)}_{0,1,1,-2}(0)-\dfrac{\tilde{V}^{(a,a,a,a)}_{1,0,1,-2}(0)}{2\pi},
\end{equation}
\begin{multline}
\label{eq:C16}
\delta_{\alpha\alpha}^{(e-e)}=\sum_{n,i}\nu_{n}^{(i)}\left(\dfrac{\tilde{V}^{(a,i,i,a)}_{1,n,n,1}(0)}{2\pi}-\dfrac{\tilde{V}^{(a,i,i,a)}_{0,n,n,0}(0)}{2\pi}
-\tilde{E}^{(a,i,a,i)}_{1,n,1,n}(0)+\tilde{E}^{(a,i,a,i)}_{0,n,0,n}(0)\right)+\\
+\left(\nu_{0}^{(a)}-\nu_{1}^{(a)}\right)\left(\tilde{E}^{(a,a,a,a)}_{0,1,1,0}(0)-\dfrac{\tilde{V}^{(a,a,a,a)}_{1,0,1,0}(0)}{2\pi}  \right),
\end{multline}
and
\begin{multline}
\label{eq:C17}
\delta_{\alpha'\alpha'}^{(e-e)}=\sum_{n,i}\nu_{n}^{(i)}\left(\dfrac{\tilde{V}^{(a,i,i,a)}_{1,n,n,1}(0)}{2\pi}-\dfrac{\tilde{V}^{(a,i,i,a)}_{-2,n,n,-2}(0)}{2\pi}
-\tilde{E}^{(a,i,a,i)}_{1,n,1,n}(0)+\tilde{E}^{(a,i,a,i)}_{-2,n,-2,n}(0)\right)+\\
+\left(\nu_{-2}^{(a)}-\nu_{1}^{(a)}\right)\left(\tilde{E}^{(a,a,a,a)}_{-2,1,1,-2}(0)-\dfrac{\tilde{V}^{(a,a,a,a)}_{1,-2,1,-2}(0)}{2\pi}  \right).
\end{multline}
Note that the summation in Eqs.~(\ref{eq:C16}) and (\ref{eq:C17}) should be performed \emph{over all LLs} in HgTe QW.

Similarly, the effective Hamiltonian describing the hybridization of $\beta$ and $\beta'$ LL transitions is written as
\begin{equation}
\label{eq:C18}
\hat{H}_{\mathrm{ME}\beta\beta'}=
\begin{pmatrix}
\hbar\omega_{\beta}^{(0)}+\delta_{\beta\beta}^{(e-e)} & \left\{\nu_{-2}^{(a)}-\nu_{-1}^{(a)}\right\}\Delta_{\beta\beta'}^{(e-e)}/2 \\
\left\{\nu_{0}^{(a)}-\nu_{-1}^{(a)}\right\}\Delta_{\beta\beta'}^{(e-e)}/2 & \hbar\omega_{\beta'}^{(0)}+\delta_{\beta'\beta'}^{(e-e)}
\end{pmatrix},
\end{equation}
where $\hbar\omega_{\beta}^{(0)}=E_{-1}^{(a)}-E_{-2}^{(a)}$ and $\hbar\omega_{\beta'}^{(0)}=E_{-1}^{(a)}-E_{0}^{(a)}$, while
\begin{equation}
\label{eq:C19}
\dfrac{\Delta_{\beta\beta'}^{(e-e)}}{2}=\tilde{E}^{(a,a,a,a)}_{-2,-1,-1,0}(0)-\dfrac{\tilde{V}^{(a,a,a,a)}_{-1,-2,-1,0}(0)}{2\pi},
\end{equation}
\begin{multline}
\label{eq:C20}
\delta_{\beta\beta}^{(e-e)}=\sum_{n,i}\nu_{n}^{(i)}\left(\dfrac{\tilde{V}^{(a,i,i,a)}_{-1,n,n,-1}(0)}{2\pi}-\dfrac{\tilde{V}^{(a,i,i,a)}_{-2,n,n,-2}(0)}{2\pi}
-\tilde{E}^{(a,i,a,i)}_{-1,n,-1,n}(0)+\tilde{E}^{(a,i,a,i)}_{-2,n,-2,n}(0)\right)+\\
+\left(\nu_{-2}^{(a)}-\nu_{-1}^{(a)}\right)\left(\tilde{E}^{(a,a,a,a)}_{-2,-1,-1,-2}(0)-\dfrac{\tilde{V}^{(a,a,a,a)}_{-1,-2,-1,-2}(0)}{2\pi}  \right),
\end{multline}
and
\begin{multline}
\label{eq:C21}
\delta_{\beta'\beta'}^{(e-e)}=\sum_{n,i}\nu_{n}^{(i)}\left(\dfrac{\tilde{V}^{(a,i,i,a)}_{-1,n,n,-1}(0)}{2\pi}-\dfrac{\tilde{V}^{(a,i,i,a)}_{0,n,n,0}(0)}{2\pi}
-\tilde{E}^{(a,i,a,i)}_{-1,n,-1,n}(0)+\tilde{E}^{(a,i,a,i)}_{0,n,0,n}(0)\right)+\\
+\left(\nu_{0}^{(a)}-\nu_{-1}^{(a)}\right)\left(\tilde{E}^{(a,a,a,a)}_{0,-1,-1,0}(0)-\dfrac{\tilde{V}^{(a,a,a,a)}_{-1,0,-1,0}(0)}{2\pi}  \right).
\end{multline}
In the main text, when discussing the hybridization of $\alpha$, $\alpha'$, $\beta$ and $\beta'$ LL transitions near the critical field $B_c$, the index ``$a$'' is omitted for brevity.

\subsection{E. Line amplitudes for the $\alpha$ and $\alpha'$ transitions within the magnetic-exciton Hamiltonian}
In this section, we present a detailed derivation of the optical line amplitudes for the $\alpha$ and $\alpha'$ LL transitions within the ME formalism. The analysis is performed in the long-wavelength limit ($\mathbf{k}\rightarrow 0$) and within the two-level approximation for the $(\alpha, \alpha')$ transition pair. The consideration for the $\beta$ and $\beta'$ LL transitions can be performed in the similar way.

Following Eq.~(\ref{eq:basisA}), the relevant ME basis states are defined as
\begin{equation}
|\alpha\rangle=A_{1,0,a,a}^{+}(\vec{k})\,|0\rangle, \qquad
|\alpha'\rangle=A_{1,-2,a,a}^{+}(\vec{k})\,|0\rangle,
\label{eq:D1}
\end{equation}
where $|0\rangle$ denotes the many-particle ground state without MEs.

Due to the non-canonical commutation relations of the ME operators in Eq.~(\ref{eq:C4}), the basis states (\ref{eq:D1}) are not normalized. In the long-wavelength limit, one finds
\begin{equation}
\langle 0|A_{n,n',i,i'}(0)A^{+}_{n,n',i,i'}(0)|0\rangle =\nu_{n'}^{(i')}-\nu_{n}^{(i)}.
\label{eq:D5}
\end{equation}
Therefore, the normalized ME basis states for the $\alpha$ and $\alpha'$ LL transitions are represented by
\begin{equation}
\label{eq:D7}
|\tilde\alpha\rangle=\dfrac{1}{\sqrt{\mathcal N_\alpha}}|\alpha\rangle,\qquad
|\tilde\alpha'\rangle=\dfrac{1}{\sqrt{\mathcal N_{\alpha'}}}|\alpha'\rangle ,
\end{equation}
where 
\begin{equation}
\mathcal N_\alpha=\left|\nu_0-\nu_1\right|, \qquad
\mathcal N_{\alpha'}=\left|\nu_{-2}-\nu_1\right|.
\label{eq:D6}
\end{equation}
Note that normalization of the ME basis states does not change the eigenvalues of $\hat{H}_{\mathrm{ME}\alpha\alpha'}$ in Eq.~(\ref{eq:C14}), but only allows it to be reduced to a Hermitian form:
\begin{equation}
\label{eq:D8}
\tilde H_{\alpha\alpha'}=
\begin{pmatrix}
\hbar\omega_{\alpha}^{(0)}+\delta_{\alpha\alpha}^{(e-e)} & g_{\alpha\alpha'} \\
g_{\alpha\alpha'} & \hbar\omega_{\alpha'}^{(0)}+\delta_{\alpha'\alpha'}^{(e-e)}
\end{pmatrix},
\end{equation}
where
\begin{equation}
\label{eq:D8b}
g_{\alpha\alpha'}=\frac{\Delta_{\alpha\alpha'}^{(e-e)}}{2}
\sqrt{\mathcal N_\alpha\mathcal N_{\alpha'}}.
\end{equation}
The latter is nothing more than half of the hybridization energy $\Delta$, whose values are extracted from the analysis of experimental data in the main text. Defining
\begin{equation}
\delta_{\alpha\alpha'}=\dfrac{\hbar\omega_{\alpha}^{(0)}-\hbar\omega_{\alpha'}^{(0)}+\delta_{\alpha\alpha}^{(e-e)}
-\delta_{\alpha'\alpha'}^{(e-e)}}{2},
\label{eq:D9}
\end{equation}
the eigenvalues of $\tilde H_{\alpha\alpha'}$ are
\begin{equation}
\label{eq:D10}
E_{\pm}=\dfrac{\hbar\omega_{\alpha}^{(0)}+\hbar\omega_{\alpha'}^{(0)}+\delta_{\alpha\alpha}^{(e-e)}
+\delta_{\alpha'\alpha'}^{(e-e)}}{2}
\pm
\sqrt{\delta_{\alpha\alpha'}^2+g_{\alpha\alpha'}^{\,2}}.
\end{equation}

The corresponding eigenstates are written as
\begin{equation}
|+\rangle=\cos\gamma\,|\tilde\alpha\rangle+\sin\gamma\,|\tilde\alpha'\rangle,
\qquad
|-\rangle=-\sin\gamma\,|\tilde\alpha\rangle+\cos\gamma\,|\tilde\alpha'\rangle ,
\label{eq:D11}
\end{equation}
where the mixing angle $\gamma$ is defined through
\begin{equation}
\tan 2\gamma=\frac{g_{\alpha\alpha'}}{\delta_{\alpha\alpha'}},
\label{eq:D12}
\end{equation}
which results in
\begin{equation}
\label{eq:D13}
\sin\gamma=\sqrt{\dfrac{1}{2}\left(1-\dfrac{\delta_{\alpha\alpha'}}{\sqrt{\delta_{\alpha\alpha'}^2+g_{\alpha\alpha'}^{\,2}}}\right)},
\qquad
\cos\gamma=\sqrt{\dfrac{1}{2}\left(1+\dfrac{\delta_{\alpha\alpha'}}{\sqrt{\delta_{\alpha\alpha'}^2+g_{\alpha\alpha'}^{\,2}}}\right)}.
\end{equation}

In the dipole approximation, the interaction with the electromagnetic field is
\begin{equation}
\hat H_{\mathrm{em}}=-e\,\mathbf E\cdot\hat{\mathbf r}.
\label{eq:D14}
\end{equation}
Projecting the coordinate operator onto the $(\alpha,\alpha')$ subspace in the long-wavelength limit yields
\begin{equation}
\hat D
=d_\alpha\,A_{1,0,a,a}^{+}(0)+d_{\alpha'}\,A_{1,-2,a,a}^{+}(0)+\mathrm{h.c.},
\label{eq:D15}
\end{equation}
where $d_\alpha$ and $d_{\alpha'}$ are the dipole matrix elements determined by the single-particle LL wave functions.

The optical matrix element for excitation between the $|\tilde\alpha\rangle$ basis state and the ground state $|0\rangle$ is written as
\begin{equation}
\label{eq:D16x}
\langle\tilde\alpha|\hat D|0\rangle=\dfrac{1}{\sqrt{\mathcal N_\alpha}}\langle\alpha|\hat D|0\rangle=
\dfrac{d_\alpha}{\sqrt{\mathcal N_\alpha}}\langle\alpha|A_{1,0,a,a}^{+}(0)|0\rangle=\dfrac{d_\alpha}{\sqrt{\mathcal N_\alpha}}\langle\alpha|\alpha\rangle=
d_\alpha\sqrt{\mathcal N_\alpha}.
\end{equation}
Similarly, we have
\begin{equation}
\label{eq:D16}
\langle\tilde\alpha'|\hat D|0\rangle=d_{\alpha'}\sqrt{\mathcal N_{\alpha'}}.
\end{equation}

The line amplitudes are defined by the optical intensity of a given hybridized ME mode, which is proportional to
\begin{equation}
I_{\pm}\propto |\langle \pm|\hat D|0\rangle|^2 .
\label{eq:D18}
\end{equation}
Using Eq.~(\ref{eq:D11}), one obtains
\begin{align}
I_{+} &\propto \left|\sqrt{\mathcal N_\alpha}\cos\gamma\,d_\alpha+
\sqrt{\mathcal N_{\alpha'}}\sin\gamma\,d_{\alpha'}\right|^2,
\notag\\
I_{-} &\propto \left|-\sqrt{\mathcal N_\alpha}\sin\gamma\,d_\alpha+
\sqrt{\mathcal N_{\alpha'}}\cos\gamma\,d_{\alpha'}\right|^2 .
\label{eq:D20}
\end{align}

Expanding these expressions yields
\begin{align}
I_{+} &\propto
{\mathcal N_\alpha}\cos^2\gamma\,|d_\alpha|^2
+{\mathcal N_{\alpha'}}\sin^2\gamma\,|d_{\alpha'}|^2
+\sqrt{\mathcal N_\alpha}\sqrt{\mathcal N_{\alpha'}}\sin2\gamma\,\mathrm{Re}\!\left(d_\alpha d_{\alpha'}^{*}\right),
\notag\\
I_{-} &\propto
{\mathcal N_\alpha}\sin^2\gamma\,|d_\alpha|^2
+{\mathcal N_{\alpha'}}\cos^2\gamma\,|d_{\alpha'}|^2
-\sqrt{\mathcal N_\alpha}\sqrt{\mathcal N_{\alpha'}}\sin2\gamma\,\mathrm{Re}\!\left(d_\alpha d_{\alpha'}^{*}\right).
\label{eq:D22}
\end{align}
Importantly, within the two-level approximation, the total oscillator strength is conserved:
\begin{equation}
I_{+}+I_{-}\propto {\mathcal N_\alpha}|d_\alpha|^2+{\mathcal N_{\alpha'}}|d_{\alpha'}|^2.
\label{eq:D23}
\end{equation}

It is important to note that if one of the transitions is optically ``dark'', i.e., characterized by a small dipole matrix element, as is the case for the $\alpha'$ LL transition with $|d_{\alpha'}|\!\ll\!|d_\alpha|$, its hybridization with a ``bright'' transition can render it optically visible. In particular, hybridization with the bright $\alpha$ transition leads to a finite oscillator strength of the $\alpha'$-derived ME mode in the magnetic-field range where the detuning $\delta_{\alpha\alpha'}$ is small. In the limit $|d_{\alpha'}|\rightarrow 0$, Eq.~(\ref{eq:D22}) reduces to
\begin{align}
I_{+} &\propto
\frac{\mathcal N_\alpha |d_\alpha|^2}{2}
\left(1+\frac{\delta_{\alpha\alpha'}}{\sqrt{\delta_{\alpha\alpha'}^2+g_{\alpha\alpha'}^{\,2}}}\right),
\notag\\
I_{-} &\propto
\frac{\mathcal N_\alpha |d_\alpha|^2}{2}
\left(1-\frac{\delta_{\alpha\alpha'}}{\sqrt{\delta_{\alpha\alpha'}^2+g_{\alpha\alpha'}^{\,2}}}\right),
\label{eq:D24}
\end{align}
where $\delta_{\alpha\alpha'}=0$ corresponds to the critical magnetic field $B_c$, extracted from the analysis of the transition-energy difference using Eq.~(2) in the main text. Equation~(\ref{eq:D24}) explicitly demonstrates that the visibility of the $\alpha'$ transition is governed not only by the hybridization energy $2g_{\alpha\alpha'}$, but also by the LL filling factors entering $\mathcal N_\alpha$. 

Finally, we note that Eq~(\ref{eq:D24}) can also be used for the extraction of the parameters $M$, $B_c$, and $\Delta$ from the relative amplitudes of the absorption lines.
Indeed, 
\begin{equation}
\label{eq:Eshort}
\left(\frac{I_+ - I_-}{I_+ + I_-}\right)^2
=
\dfrac{\delta_{\alpha\alpha'}^2}
{\delta_{\alpha\alpha'}^2+g_{\alpha\alpha'}^{\,2}}.
\end{equation}
Taking into account that
\begin{equation}
\label{eq:D25}
\delta_{\alpha\alpha'}(B)=M\!\left(1-\dfrac{B}{B_c}\right),
\end{equation}
we finally have
\begin{equation}
\label{eq:D26}
\left(\dfrac{I_+ - I_-}{I_+ + I_-}\right)^2
=
\dfrac{(2M)^2\left(1-\dfrac{B}{B_c}\right)^2}
{(2M)^2\left(1-\dfrac{B}{B_c}\right)^2+\Delta^{\,2}}.
\end{equation}
In practice, however, a quantitative extraction of the parameters $M$, $B_c$, and $\Delta$ from the absorption-line amplitudes alone is hindered by several additional effects not directly related to the ME hybridization mechanism. First, the experimental determination of the relative line amplitudes requires a reliable decomposition of partially overlapping absorption lines, which becomes increasingly uncertain in the vicinity of $B_c$, where the hybridized modes approach each other and their linewidths may be comparable to the hybridization gap. Second, the nominally dark transition generally possesses a finite, although small, dipole matrix element, so that the condition $|d_{\alpha'}|\!\ll\!|d_\alpha|$ is only approximately satisfied, leading to additional interference contributions to the line intensities. Finally, temperature-dependent broadening, background absorption, and variations in the frequency-dependent response of the detector and optical setup can affect the absolute and relative line amplitudes without modifying the hybridization energies. As a result, while the amplitude analysis provides a useful qualitative consistency check for the ME picture, the extraction of $M$, $B_c$, and $\Delta$ is significantly more robust when based on the transition energies.


%

\end{document}